\documentclass[reprint,nofootinbib,onecolumn,preprintnumbers]{revtex4}
\usepackage{amssymb}
\usepackage{color}

\usepackage{amsthm}
\usepackage{graphicx}
\usepackage{dcolumn}
\usepackage{bm}
\usepackage{subfigure}
\usepackage{amssymb}
\usepackage{verbatim}
\usepackage{amscd}
\usepackage{amssymb}
\usepackage{amsmath, amsfonts}
\usepackage{setspace}
\usepackage[]{graphicx}        
\usepackage{amsthm}
\usepackage{enumerate}
\newcommand{\tb}[1]{\textbf{#1}}

\theoremstyle{plain}

\theoremstyle{plain}

\theoremstyle{plain}
 
\theoremstyle{plain}

\theoremstyle{remark}

\theoremstyle{conjecture}

\theoremstyle{observation}

\theoremstyle{definition}

\theoremstyle{corollary}

\theoremstyle{definition}

\theoremstyle{definition}

\theoremstyle{assumption}

\theoremstyle{definition}

\theoremstyle{problem}

\theoremstyle{fact}

\begin{document}

\title{Topological phases with generalized global symmetries}
\author{Beni Yoshida}
\affiliation{Walter Burke Institute for Theoretical Physics and Institute for Quantum Information \& Matter, California Institute of Technology, Pasadena, California 91125, USA
}
\affiliation{Kavli Institute for Theoretical Physics, University of California, Santa Barbara, CA 93106, USA}

\preprint{NSF-KITP-15-095}

\date{\today}

\begin{abstract}
We present simple lattice realizations of symmetry-protected topological (SPT) phases with $q$-form global symmetries where charged excitations have $q$ spatial dimensions. Specifically, we construct $d$ space-dimensional models supported on a $(d+1)$-colorable graph by using a family of unitary phase gates, known as multi-qubit control-$Z$ gates in quantum information community. In our construction, charged excitations of different dimensionality may coexist and form a short-range entangled state which is protected by symmetry operators of different dimensionality. Non-triviality of proposed models, in a sense of quantum circuit complexity, is confirmed by studying protected boundary modes, gauged models and corresponding gapped domain walls. We also comment on applications of our construction to quantum error-correcting codes, and discuss corresponding fault-tolerant logical gates. 
\end{abstract}

\maketitle

\section{Introduction}

The study of symmetry-protected topological (SPT) phases has attracted a considerable amount of attention~\cite{Dijkgraaf90, Propitius95, Kitaev09, Pollmann12, Chen11, Chen11b, Lu12, Schuch11, Fidkowski11, Walker11, Levin12, Chen13, Keyserlingk13, Hu13, Vishwanath13, Metlitski13, Wen13, Wang13, Burnell14, Gu14, Kapustin14}. Recently, generalizations of SPT phases with higher-form symmetry have been discussed~\cite{Baez04, Baez10, Kapustin13, Kapustin14c, Gaiotto15}. Ordinary SPT phases are discussed in the presence of a global $0$-form symmetry operator of the on-site form:
\begin{align}
U^{(g)} = \bigotimes_{j} U_{j}^{(g)}
\end{align}
where $g\in G$ is an element of the symmetry group $G$ and $j$ represents a lattice site. The symmetry operator imposes a conservation law where charged excitations are point-like objects. A $q$-form global symmetry can be imposed by an operator of the form $U^{(g)}(\mathcal{M})$ which acts on a closed codimension-$q$ manifold $\mathcal{M}$ (codimension $q+1$ for a space-time). In such a theory, charged excitations have $q$ space dimensions and symmetry operators impose conservation laws on higher-dimensional charged objects. There have been several pioneering works in this direction~\cite{Baez10,Kapustin13, Kapustin14c, Gaiotto15}. Namely the work by Kapustin and Thorngren proposes a family of lattice realizations by replacing a finite group by a finite 2-group~\cite{Kapustin13}.

The goal of this paper is to present simple lattice realizations of bosonic SPT phases with higher-form global symmetry and discuss their quantum information application. Bosonic SPT phases are often constructed by using mathematical machinery such as group cohomology/cobordism and their physical properties are typically explained via gauge/gravity anomalies~\cite{Chen13, Kapustin13, Kapustin14, Kapustin14b, Wen13}. In this paper, we shall restrict our attentions to rather simple realizations with $\mathbb{Z}_{2}$ symmetry and study their ``properties'' in a systematic manner. (Generalization to $\mathbb{Z}_{N}$ symmetry, or arbitrary abelian symmetry, is also possible). The proposed models have a short-range entangled unique gapped ground state on a closed manifold which is protected by higher-form global symmetry. In our construction, charged excitations of different dimensionality may coexist and form a short-range entangled state which is protected by symmetry operators of different dimensionality. For instance, to provide some insight, we mention the existence of a non-trivial $(5+1)$-dimensional model protected by $0$-form, $1$-form and $2$-form $\mathbb{Z}_{2}\otimes \mathbb{Z}_{2}\otimes \mathbb{Z}_{2}$ symmetry where charged particles, loops and membranes form a short-range entangled vacuum. Kapustin and Thorngren provided a framework for constructing SPT phases protected by $0$-form and $1$-form symmetries by using $2$-group. A  complementary feature of our work is that the model admits arbitrary $q$-form symmetries and its construction is simple. Also, some of our models do not seem to fit into the framework of the $2$-group construction despite the fact that these models possess $0$-form and $1$-form symmetries only.

Let us clarify what we mean by \emph{non-trivial} SPT phases. We will say that a wavefunction is non-trivial if its preparation from a product state requires a large-depth local quantum circuit. By a local quantum circuit, we mean a unitary operation that can be implemented by a local Hamiltonian which may be time-dependent. By a large-depth, we mean that it takes a system-size dependent time to implement the required unitary operation. Ground states of (intrinsic) topologically ordered Hamiltonians are known to be non-trivial in this sense. Non-triviality of SPT wavefunctions can be defined by restricting considerations to local quantum circuits which commute with imposed symmetries. Namely, each local component of the quantum circuit needs to commute with symmetry operators. To prove the non-triviality of the wavefunctions in the presence of symmetry, we employ three different arguments. First we consider the models with open boundaries and demonstrate that they have protected boundary modes which trivial models cannot possess. Namely, boundary mode may be gapless, spontaneous symmetry-breaking phase and/or topological phase depending on dimensionality of symmetry operators and spatial dimensions. While this type of arguments has been extensively used as simple diagnostic of non-triviality of SPT wavefunctions on the bulk, its connection to the circuit complexity of wavefunctions is rather indirect. Second we minimally couple the SPT Hamiltonians to generalized (higher-form) gauge fields. Namely, for $q$-form symmetry, the system is coupled to $(q+1)$-form gauge fields. We show that gauged versions of trivial and non-trivial SPT Hamiltonians belong to different topological phases, which implies that the original Hamiltonians (and wavefunctions) belong to different quantum phases in the presence of symmetries. This type of argument was originally proposed by Levin and Gu in the context of $0$-form SPT phases, and can be converted into a fairly rigorous argument by formulating the gauging as a duality map. (See~\cite{Beni15c} for instance). Third we shall construct a gapped domain wall in a $(d+1)$-dimensional topological phase by gauging the $d$-dimensional SPT wavefunction in $(d+1)$ space dimensions by following the idea in~\cite{Beni15}. We then show that a gapped domain wall transposes excitations in a non-trivial manner which is possible only if the underlying SPT wavefunction is non-trivial. This type of argument is perhaps non-standard in condensed matter and high energy community, but is indeed natural from quantum information theoretical viewpoint. These three characterizations show that proposed models of higher-form SPT phases are indeed non-trivial in a sense of quantum circuit complexity.

There are two key ingredients in our construction. First all the models are constructed on $d$-dimensional simplicial lattices which are $d+1$-colorable, meaning that one can assign $d+1$ distinct color labels to vertices of the lattice such that neighboring vertices have different colors. Colorable graphs have a number of useful properties in discussing lattice realizations of topological theories~\cite{Bombin06, Bombin07, Bombin15, Kubica15}. For instance, we shall see that gauged models can be defined on the same graph without modifying the lattice structure. The simplest example of colorable graphs is a two-dimensional triangular lattice where three color labels can be assigned to vertices. Second, to construct non-trivial models, we shall use unitary phase gates, known as multi-qubit control-$Z$ gates in quantum information community, which are closely related to a certain non-trivial $m$-cocycle function for $G=(\mathbb{Z}_{2})^{\otimes m}$~\cite{Kubica15b, Beni15}. Multi-qubit control-$Z$ gates are of particular importance in the context of fault-tolerant quantum computation since they are outside of the Clifford group when involving three or more qubits~\cite{Gottesman99}.  

The overall construction is summarized as follows. Let $a_{1},a_{2},\ldots, a_{d+1}$ be $d+1$ distinct color labels. We think of splitting color labels into several groups. We then place qubits on $q$-simplexes according to the splitting. For instance, for $d=5$, one may have the following splitting:
\begin{align}
a_{1}| a_{2} a_{3} | a_{4} a_{5}a_{6}
\end{align}
and qubits are placed on $0$-simplexes of color $a_{1}$, $1$-simplexes of color $a_{2}a_{3}$ and $2$-simplexes of color $a_{4}a_{5}a_{6}$. We then apply multi-qubit control-$Z$ gates on each $d$-simplex to obtain the non-trivial Hamiltonian from a trivial Hamiltonian. In the above example, we will obtain a five-dimensional model with $(0,1,2)$-form $\mathbb{Z}_{2}\otimes \mathbb{Z}_{2} \otimes \mathbb{Z}_{2}$ symmetry by applying three-qubit control-$Z$ gates. The system supports point-like, loop-like and membrane-like charged excitations. While our studies are limited to models with $\mathbb{Z}_{2}$ symmetries, the constructions can be generalized to systems with $\mathbb{Z}_{N}$ symmetries by using a certain generalization of control-$Z$ gates as briefly explained in Section~\ref{sec:discussion}. Our construction can be viewed as a special realization of hyper-graph states recently proposed in~\cite{Guhne14}, and thus, applications to measurement-based quantum computations may be an interesting future problem. 

Classification of topological phases of matter is a problem of fundamental and practical importance, bridging condensed matter physics and quantum information science. At a formal level, lattice models of topological phases of matter can be probably classified by using the framework of higher-category theory. This, however, does not mean that classification of topological order is completed since category theoretical approaches provide only a set of consistency equations, such as pentagon and hexagon equations. Solving consistency equations are rather difficult both analytically and computationally, and thus finding a non-trivial solution to consistency equations seems to be the real challenge. Our lattice models, before and after coupling to gauge fields, presumably satisfy these consistency equations of category theoretical approaches, and are beyond known theories of topological order, such as the Walker-Wang model~\cite{Walker11} and the Dijkgraaf-Witten topological gauge theories~\cite{Dijkgraaf90}. Thus, our model may serve as a stepping stone to further looking for exotic topological phases of matter, which may be of importance for quantum information processing purposes. 

Indeed, proposed models of SPT phases with higher-form symmetry have interesting quantum coding applications. In~\cite{Beni15, Beni15c}, it has been pointed out that the classification of bosonic SPT phases with $0$-form symmetry and classification of fault-tolerant logical gates in topological quantum codes are closely related. The construction of bosonic SPT phases with higher-form symmetry enables us to find (somewhat surprising) fault-tolerantly implementable logical gates. For instance, consider a four-dimensional system consisting of two copies of the $(1,3)$-toric code and one copy of the $(2,2)$-toric code. Here, the $(a,b)$-toric code refers to the $(a+b)$-dimensional toric code with $a$-dimensional Pauli-$Z$ logical operators and $b$-dimensional Pauli-$X$ logical operators. We choose these three copies of the toric code to be decoupled from each other. As we shall see, there exists a non-trivial three-dimensional SPT model which is protected by $0$-form $\mathbb{Z}_{2}\otimes \mathbb{Z}_{2}$ symmetry and $1$-form $\mathbb{Z}_{2}$ symmetry. The presence of such an SPT phase implies that one can implement a three-qubit control-$Z$ logical gate, belonging to the third-level of the Clifford hierarchy, among three copies of the four-dimensional toric code fault-tolerantly by a finite-depth local quantum circuit. This is rather surprising given the fact that the $(1,3)$-toric code and the $(2,2)$-toric code possess logical operators of different dimensionality and thus belong to different topological phases. Our approach may give a hint on how to implement multi-qubit unitary logical gates on multiple quantum error-correcting codes of different code generators.

The paper is organized as follows. Section~\ref{sec:0-form} of the paper is devoted to a brief review of bosonic SPT phases with $0$-form symmetry. While this is a throughly explored subject, we try to provide a concise, yet precise treatment of various properties of $0$-form SPT phases, such as the boundary mode, outcome of coupling to gauge fields and corresponding gapped boundaries, by using concepts from quantum information theory. Although we do not provide any new results in this section, our treatment and observations may have some novelty. In section~\ref{sec:higher-form}, we will present lattice realizations of a bosonic SPT phase with generalized global symmetry and discuss their physical properties. We demonstrate that boundary of these models exhibits spontaneous breaking of symmetry, gapless critical modes or/and topological phases. In section~\ref{sec:gate}, we comment on quantum coding implications of our results. 


\section{Topological phases with $0$-form symmetry}\label{sec:0-form}

We begin by presenting a simple recipe of how to construct certain families of bosonic SPT wavefunctions with $0$-form symmetry by using multi-qubit control-$Z$ operators. Namely, we construct wavefunctions for $d$-dimensional SPT phases with ${\mathbb{Z}_{2}}^{\otimes d+1}$ symmetry. We then study their boundary modes by finding dressed boundary operators which commute with the bulk Hamiltonian while satisfying certain commutation relations of Pauli operators. In a quantum coding language, boundary modes and dressed boundary operators can be viewed as a codeword space and logical operators of the bulk Hamiltonian respectively. This allows us to study symmetry constraints on boundary terms and discuss the boundary mode protected by symmetry. We also discuss the procedure of coupling SPT phases to gauge fields and corresponding gapped domain walls. 

\subsection{Multi-qubit control-$Z$ gates}

\begin{figure}[htb!]
\centering
\includegraphics[width=0.65\linewidth]{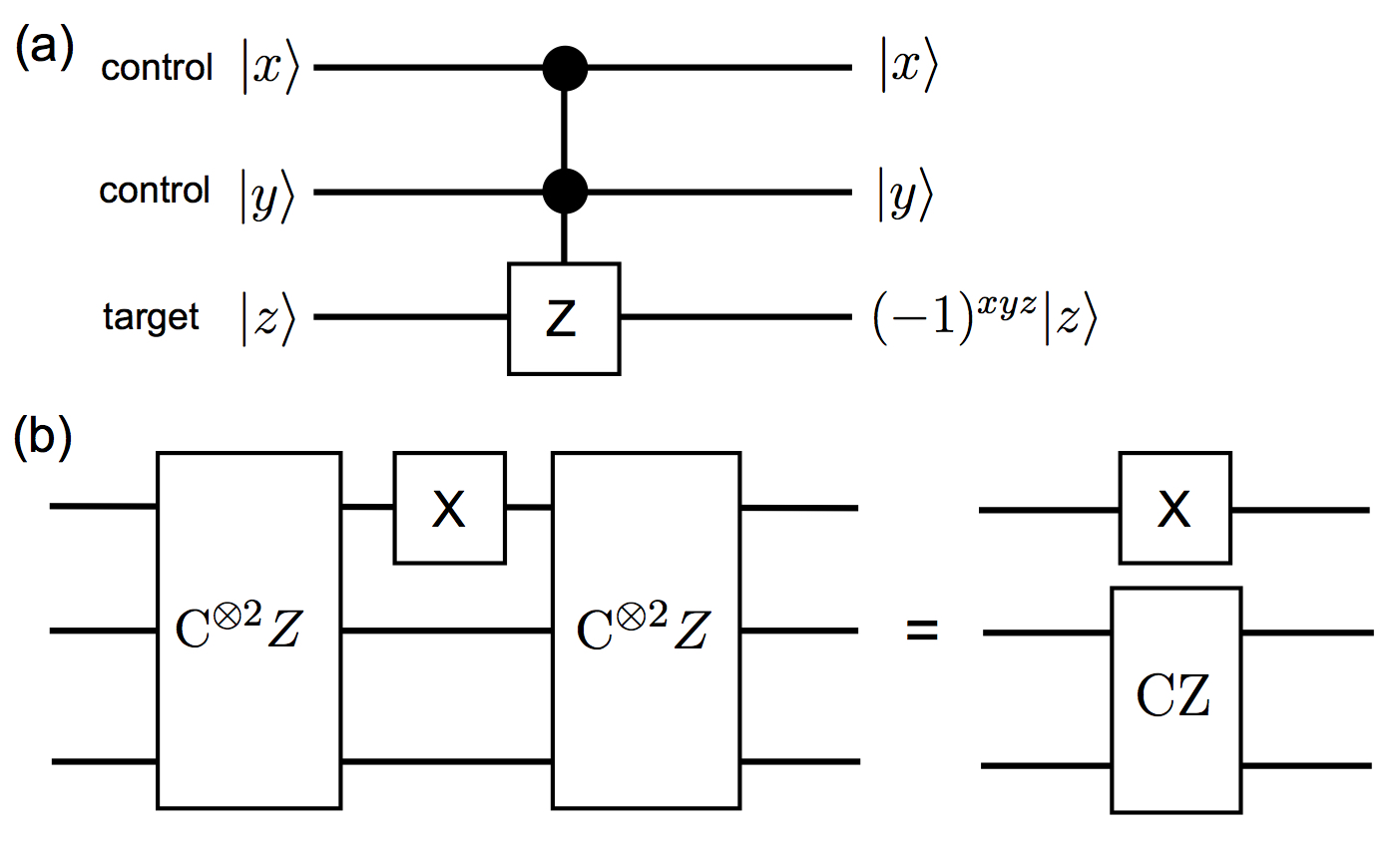}
\caption{(a) A quantum circuit representation of the three-qubit control-$Z$ gate $\mbox{C}^{\otimes 2}Z$. (b) Conjugation by the three-qubit control-$Z$ gate. Note that $(\mbox{C}^{\otimes 2}Z)^{\dagger}=\mbox{C}^{\otimes 2}Z$.
} 
\label{fig_multi_qubit_gate}
\end{figure}

We begin by presenting the definition of a multi-qubit control-$Z$ gate~\cite{Kubica15b, Beni15} (see Fig.~\ref{fig_multi_qubit_gate}(a) for its quantum circuit representation). The control-$Z$ gate, denoted by $\mbox{C}Z$, is a two-qubit phase gate acting in the computational basis as
\begin{align}
\mbox{C}Z |x,y\rangle = (-1)^{xy} |x,y\rangle \qquad x,y=0,1
\end{align}
which adds $-1$ phase if both the first and second qubits are in $|1\rangle$ state. One can generalize the control-$Z$ gate to a system of multiple qubits. The $n$-qubit control-$Z$ gate, denoted by $\mbox{C}^{\otimes n-1}Z$, acts as follows
\begin{align}
\mbox{C}^{\otimes n-1}Z |x_{1},\ldots,x_{n}\rangle = (-1)^{x_{1}\ldots x_{n}} |x_{1},\ldots,x_{n}\rangle\qquad x_{j}=0,1.
\end{align}

It is convenient to summarize how Pauli $X$ operators transform under conjugation by multi-qubit control-$Z$ operators. For a two-qubit control-$Z$ gate, one has 
\begin{align}
\mbox{C}Z (X_{1}) \mbox{C}Z= (X_{1})Z_{2} \qquad \mbox{C}Z (X_{2}) \mbox{C}Z= Z_{1}(X_{2}).
\end{align}
For a multi-qubit control-$Z$ gate, one has 
\begin{align}
\mbox{C}^{\otimes n-1}Z (X_{1}) \mbox{C}^{\otimes n-1}Z = (X_{1})\mbox{C}^{\otimes n-2}Z_{2\ldots n}.
\end{align}
Here $\mbox{C}^{\otimes n-2}Z_{2\ldots n}$ acts on the $j$th qubits ($2\leq j \leq n$). So, conjugation by $\mbox{C}^{\otimes n-1}Z$ adds ``decoration'' of $\mbox{C}^{\otimes n-2}Z$ on Pauli $X$ operators (Fig.~\ref{fig_multi_qubit_gate}(b)).

Multi-qubit control-$Z$ gates have particularly useful applications in quantum coding theory since $\mbox{C}^{\otimes n-1}Z$ belongs to the $n$-th level of the so-called Clifford hierarchy~\cite{Gottesman99, Eastin09, Bravyi13b, Bombin06, Bombin15, Pastawski15} which is an important concept in classifying fault-tolerantly implementable logical gates in topological stabilizer codes. Readers who are familiar with topological gauge theories may recognize the similarity between $\mbox{C}^{\otimes n-1}Z$ operators and a non-trivial $n$-cocycle function for $G={\mathbb{Z}_{2}}^{\otimes n}$: $\omega_{n}(g^{(1)},\ldots,g^{(n)}) = (-1)^{ g^{(1)}_{1}\ldots g^{(n)}_{n} }$ where $g^{(i)}=(g^{(i)}_{1},g^{(i)}_{2},\ldots,g^{(i)}_{n})$ and $g^{(i)}_{j}=0,1$~\cite{Propitius95, J_Wang15, J_Wang15b}. For the connection between group cohomology and the Clifford hierarchy, see~\cite{Beni15c}. 

\subsection{One-dimensional model with $ \mathbb{Z}_{2}\otimes \mathbb{Z}_{2}$ symmetry}

In this subsection, we study the one-dimensional SPT phase with $\mathbb{Z}_{2}\otimes \mathbb{Z}_{2}$ symmetry~\cite{Chen11, Schuch11}. Consider a one-dimensional chain of $2n$ qubits with periodic boundary conditions. We assign color labels $a,b$ to vertices in a bipartite manner such that odd (even) sites have color $a$ ($b$). The Hamiltonian is given by
\begin{align}
H_{1} = - \sum_{j=1}^{2n}O_{j}, \qquad O_{j}=X_{j-1}Z_{j}X_{j+1}.\label{eq:1dim_term}
\end{align}
Since $[O_{i},O_{j}]=0$, the ground state $|\psi\rangle$ satisfies $O_{j}|\psi\rangle = |\psi\rangle$ for all $j$. The Hamiltonian and the ground state have $\mathbb{Z}_{2}\otimes \mathbb{Z}_{2}$ symmetry corresponding to two symmetry operators:
\begin{align}
S^{(a)} = \prod_{j=1}^{n}X_{2j-1}, \qquad S^{(b)} = \prod_{j=1}^{n}X_{2j}.
\end{align}
where $S^{(a)},S^{(b)}$ act on vertices of color $a,b$ respectively. We can see that the Hamiltonian respects the symmetry: $[H_{1},S^{(a)}]=[H_{1},S^{(b)}]=0$, as well as the ground state: $S^{(a)}|\psi\rangle=S^{(b)}|\psi\rangle=|\psi\rangle$. To verify this, observe that $S^{(a)}=\prod_{j=1}^{n} O_{2j-1}, S^{(b)}=\prod_{j=1}^{n} O_{2j}$.

One can see that the ground state $|\psi\rangle$ is short-range entangled by considering the following finite-depth quantum circuit:
\begin{align}
U^{(0,0)} = \prod_{e\in E} \mbox{C}Z_{e}
\end{align}
where $\mbox{C}Z_{e}$ acts on two qubits on the edge $e$ and $E$ represents the set of all the edges. The superscript in $U^{(0,0)}$ indicates that we construct a model with two copies of $0$-form $\mathbb{Z}_{2}$ symmetry. One has
\begin{align}
H_{1}= U^{(0,0)}H_{0}{U^{(0,0)}}^{\dagger},\qquad H_{0}= - \sum_{j}X_{j},
\end{align}
where $|\psi\rangle = U^{(0,0)} |+\rangle^{\otimes 2n}$ and $|+\rangle := \frac{1}{\sqrt{2}}(|0\rangle + |1\rangle)$. 
However, the quantum circuit $U^{(0,0)}$ is not symmetric since each local component, $\mbox{C}Z_{e}$, does not commute with symmetry operators $S^{(a)}$ or $S^{(b)}$. (One important subtlety is that $U^{(0,0)}$ commutes with $S^{(a)}$ and $S^{(b)}$ as a whole, but one cannot implement it through local quantum gates which commute with $S^{(a)}$ and $S^{(b)}$). 
Indeed, one can verify that there is no finite-depth symmetric quantum circuit which creates $|\psi\rangle$ from a product state. See~\cite{Huang14} for instance. In this sense, we say that $|\psi\rangle$ is a non-trivial SPT wavefunction in the presence of $\mathbb{Z}_{2}\otimes \mathbb{Z}_{2}$ symmetry.

\begin{figure}[htb!]
\centering
\includegraphics[width=0.55\linewidth]{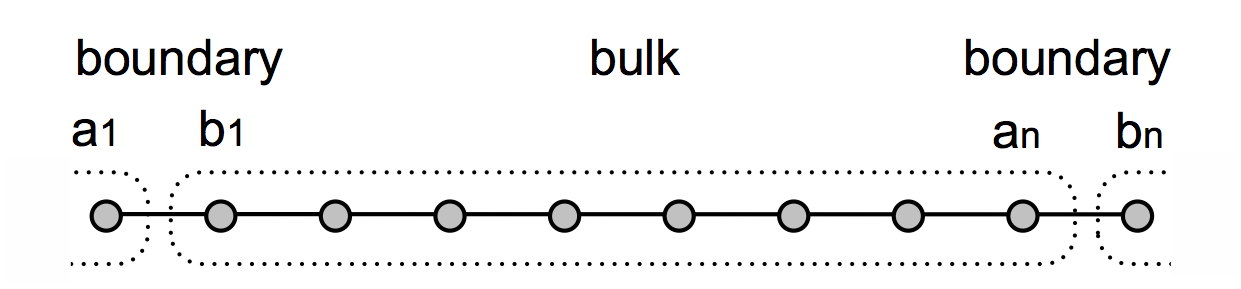}
\caption{Bulk and boundary vertices on a one-dimensional chain with boundaries.
} 
\label{fig_1dim_edge}
\end{figure}

We then study the boundary mode for the one-dimensional $\mathbb{Z}_{2}\otimes \mathbb{Z}_{2}$ SPT phase by following an approach used by Levin and Gu~\cite{Levin12}. Consider a one-dimensional chain of $2n$ spins with boundaries. Let $\tb{bulk}$ and $\tb{boundary}$ be sets of vertices in the bulk and on the boundary respectively as shown in Fig.~\ref{fig_1dim_edge}. We labeled qubits by $a_{1},b_{1},\ldots,a_{n},b_{n}$, and $\tb{boundary}=\{a_{1},b_{n}\}$. One can write a generic form of the Hamiltonian as follows:
\begin{align}
H = H_{boundary} + H_{bulk}\qquad H_{bulk}= - J \sum_{j \in \tb{bulk} } O_{j}
\end{align}
where $H_{boundary}$ involves terms localized near the boundaries and $O_{j}$ are given by Eq.~(\ref{eq:1dim_term}). We assume that the interaction strength $J>0$ is sufficiently large so that one can restrict attentions to the low-energy subspace $\mathcal{C}$:
\begin{align}
\mathcal{C} = \left\{ |\psi\rangle : O_{j}|\psi\rangle = |\psi\rangle \qquad j \in \tb{bulk} \right\}.
\end{align}
The low-energy subspace $\mathcal{C}$ is a $4$-dimensional space. In the large $J$ limit, one can assume that the boundary terms commute with the bulk Hamiltonian:
\begin{align}
[H_{boundary},H_{bulk}]=0.
\end{align}
Later, we will discuss the cases where $H_{boundary}$ does not commute with $H_{bulk}$. 

We hope to find operators which characterize the boundary mode associated with the low-energy subspace $\mathcal{C}$. Ordinary boundary Pauli operators, $X_{a_{1}},Z_{a_{1}},X_{b_{n}},Z_{b_{n}}$ on the boundary, do not commute with $H_{bulk}$, and thus are not appropriate operators to describe physics of the boundary mode. One needs to find a complete set of Pauli operators which act non-trivially inside $\mathcal{C}$, but preserve $\mathcal{C}$. Let us consider the quantum circuit $U^{(0,0)}$ which is truncated at the boundary: $U^{(0,0)}= \prod_{e\in E}\mbox{C}Z_{e}=\mbox{C}Z_{a_{1}b_{1}}\mbox{C}Z_{b_{1}a_{2}}\ldots \mbox{C}Z_{b_{n-1}a_{n}}\mbox{C}Z_{a_{n}b_{n}}$. The following dressed boundary operators play the role of Pauli operators:
\begin{equation}
\begin{split}
&\overline{X}_{a_{1}} = U^{(0,0)} X_{a_{1}} {U^{(0,0)}}^{\dagger} =X_{a_{1}}Z_{b_{1}},\qquad
\overline{Z}_{a_{1}} = U^{(0,0)} Z_{a_{1}} {U^{(0,0)}}^{\dagger} =Z_{a_{1}} \\
&\overline{X}_{b_{n}} = U^{(0,0)} X_{b_{n}} {U^{(0,0)}}^{\dagger} =Z_{b_{n}}X_{b_{n}},\qquad
\overline{Z}_{b_{n}} = U^{(0,0)} Z_{b_{n}} {U^{(0,0)}}^{\dagger} = Z_{b_{n}}.
\end{split}
\end{equation}
One can verify that these Pauli operators commute with $O_{j}$ for $j\in \tb{bulk}$. As such, $\overline{X}_{a_{1}},\overline{Z}_{a_{1}}$ characterize the left boundary mode and $\overline{X}_{b_{n}},\overline{Z}_{b_{n}}$ characterize the right boundary mode. In the language of quantum coding theory, one may view $\mathcal{C}$ as a codeword space, and $\overline{X}_{a_{1}}, \overline{Z}_{a_{1}}, \overline{X}_{b_{n}},\overline{Z}_{b_{n}}$ as logical operators acting non-trivially inside $\mathcal{C}$. In this picture, boundary modes are logical qubits encoded in the codeword space $\mathcal{C}$. These boundary operators are ``dressed'' in a sense that they involve Pauli operators on the bulk. 

While dressed boundary operators commute with the bulk terms, they are not symmetric under $S^{(a)}$ and $S^{(b)}$. Indeed, dressed boundary operators are transformed under conjugation by $S^{(a)}$ and $S^{(b)}$ as follows:
\begin{equation}
\begin{split}
S^{(a)}:\overline{X}_{a_{1}} \rightarrow  \overline{X}_{a_{1}} \qquad \overline{Z}_{a_{1}} \rightarrow - \overline{Z}_{a_{1}} \qquad 
\overline{X}_{b_{n}} \rightarrow - \overline{X}_{b_{n}} \qquad \overline{Z}_{b_{n}} \rightarrow  \overline{Z}_{b_{n}} \\
S^{(b)}:\overline{X}_{a_{1}} \rightarrow -\overline{X}_{a_{a}} \qquad \overline{Z}_{a_{1}} \rightarrow  \overline{Z}_{a_{1}} \qquad 
\overline{X}_{b_{n}} \rightarrow \overline{X}_{b_{n}} \qquad \overline{Z}_{b_{n}} \rightarrow - \overline{Z}_{b_{n}}.
\end{split}
\end{equation}
From the above relations, one can deduce the action of symmetry operators inside $\mathcal{C}$ as follows:
\begin{align}
S^{(a)} \sim \overline{X}_{a_{1}}\otimes \overline{Z}_{b_{n}},
\qquad S^{(b)} \sim  \overline{Z}_{a_{1}}\otimes \overline{X}_{b_{n}}.
\end{align}
The notation ``$\sim$'' implies that two operators act in an identical manner inside the low-energy subspace $\mathcal{C}$. We can then ask what kinds of boundary terms are allowed under this symmetry. Observe that a term on the left boundary needs to commute with $\overline{X}_{a_{1}}$ and $\overline{Z}_{a_{1}}$, implying that there is no term that can be added on the left boundary. A similar argument holds for the right boundary. The only possible terms for $H_{boundary}$ are $S^{(a)}$ and $S^{(b)}$ which are highly non-local. Therefore, the degeneracy on the edges cannot be lifted. One can consider boundary terms $H_{boundary}$ which do not commute with $H_{bulk}$ too. In such cases, perturbative analysis implies that non-trivial coupling between four degenerate ground states appear only in the $O(n)$-th order perturbative expansion which is exponentially suppressed. Therefore, one expects that the energy splitting among four low-energy states is exponentially small with respect to the system size $n$. The conclusion is that four degenerate boundary states are protected by symmetry and the degeneracy cannot be lifted by small perturbations which respect the imposed symmetry.  

\subsection{Two-dimensional model with $\mathbb{Z}_{2}\otimes \mathbb{Z}_{2}\otimes \mathbb{Z}_{2}$ symmetry}

\begin{figure}[htb!]
\centering
\includegraphics[width=0.7\linewidth]{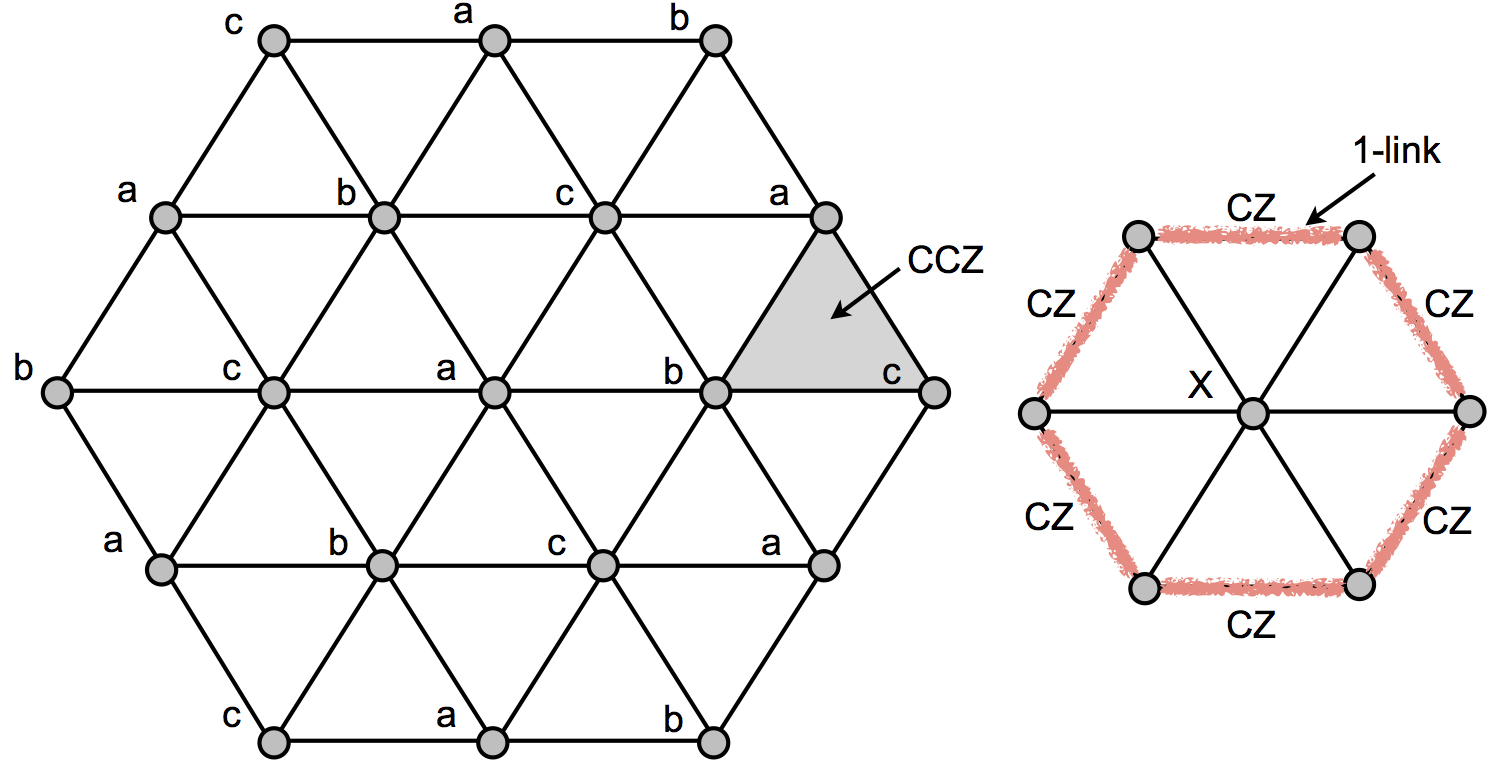}
\caption{Two-dimensional SPT phase with $\mathbb{Z}_{2}\otimes \mathbb{Z}_{2}\otimes \mathbb{Z}_{2}$ symmetry. Each of $\mathbb{Z}_{2}$ symmetry operators acts on qubits of distinct colors. The quantum circuit for this system consists of $\mbox{CCZ}$ acting on each triangle. The figure on the right shows an interaction term which is a Pauli $X$ operator at a vertex $v$ decorated by $\mbox{C}Z$ operators acting on $1$-links of $v$. 
} 
\label{fig_2dimSPT}
\end{figure}

In this subsection, we shall study the two-dimensional SPT phase with $\mathbb{Z}_{2}\otimes \mathbb{Z}_{2}\otimes \mathbb{Z}_{2}$ symmetry. Consider a triangular lattice as depicted in Fig.~\ref{fig_2dimSPT}, which is three-colorable in a sense that one can assign color labels $a,b,c$ to vertices in such a way that neighboring vertices have different color labels. Qubits are placed on vertices of the lattice. Consider the trivial Hamiltonian $H_{0}=-\sum_{v\in V}X_{v}$ where $V$ represents the set of all the vertices. We shall apply the following finite-depth quantum circuit, consisting of $\mbox{CC}Z$ operators acting on each triple of qubits contained inside a triangle:
\begin{align}
U^{(0,0,0)} := \prod_{(i,j,k)\in\Delta}\mbox{CC}Z_{i,j,k}
\end{align}
where $\Delta$ represents the set of all the triangles. The resulting non-trivial Hamiltonian can be written as 
\begin{align}
H_{1} = U^{(0,0,0)}H_{0}{U^{(0,0,0)}}^{\dagger} = - \sum_{v\in V} O_{v},\qquad
O_{v} = X_{v} \prod_{e\in \tb{$1$-link}(v)} \mbox{C}Z_{e}
\end{align}
where $\tb{$1$-link}(v)$ represents the set of all the $1$-links of a vertex $v$ (Fig.~\ref{fig_2dimSPT}). (A $1$-link of $v$ forms a $2$-simplex by adding $v$. For a precise definition of $1$-link, see~\cite{Kubica15}). Thus, interaction terms are Pauli $X$ operators decorated by $\mbox{C}Z$ operators.

This Hamiltonian has $\mathbb{Z}_{2}\otimes \mathbb{Z}_{2}\otimes \mathbb{Z}_{2}$ symmetry with respect to the following triple of $\mathbb{Z}_{2}$ symmetry operators:
\begin{align}
S^{(a)}=\bigotimes_{v \in V^{(a)}}X_{v}, \qquad S^{(b)}=\bigotimes_{v \in V^{(b)}}X_{v},\qquad S^{(c)}=\bigotimes_{v\in V^{(c)}}X_{v}
\end{align}
where $V^{(a)},V^{(b)},V^{(c)}$ represent the sets of all the vertices of color $a,b,c$ respectively. That is, Pauli $X$ operators act on qubits with distinct colors. Direct calculation shows that the Hamiltonian respects the $\mathbb{Z}_{2}\otimes \mathbb{Z}_{2}\otimes \mathbb{Z}_{2}$ symmetry: $[H_{1},S^{(a)}]=[H_{1},S^{(b)}]=[H_{1},S^{(c)}]=0$. Also, the ground state $|\psi\rangle$ is symmetric since $\prod_{v\in V^{(a)}} O_{v} = S^{(a)}, \prod_{v\in V^{(b)}} O_{v} = S^{(b)}, \prod_{v\in V^{(c)}} O_{v} = S^{(c)}$. To obtain this, recall that $(\mbox{C}Z)^{2}=I$. As such, the Hamiltonian $H_{1}$ possesses $\mathbb{Z}_{2}\otimes \mathbb{Z}_{2}\otimes \mathbb{Z}_{2}$ symmetry. Under $\mathbb{Z}_{2}\otimes \mathbb{Z}_{2}\otimes \mathbb{Z}_{2}$ symmetry, there are $128=2^{7}$ different SPT phases with $7$ distinct generators which can be sorted into three types, called type-I, type-II and type-III. Upon gauging, type-I and type-II models are dual to abelian quantum double model with semionic statistics while type-III model is dual to the non-abelian $D_{4}$ quantum double model~\cite{Propitius95, Hung14, J_Wang15, J_Wang15b}. It has been shown that the aforementioned model corresponds to the SPT phase which contains the type-III cocycle function~\cite{Beni15c}. 

\begin{figure}[htb!]
\centering
\includegraphics[width=0.50\linewidth]{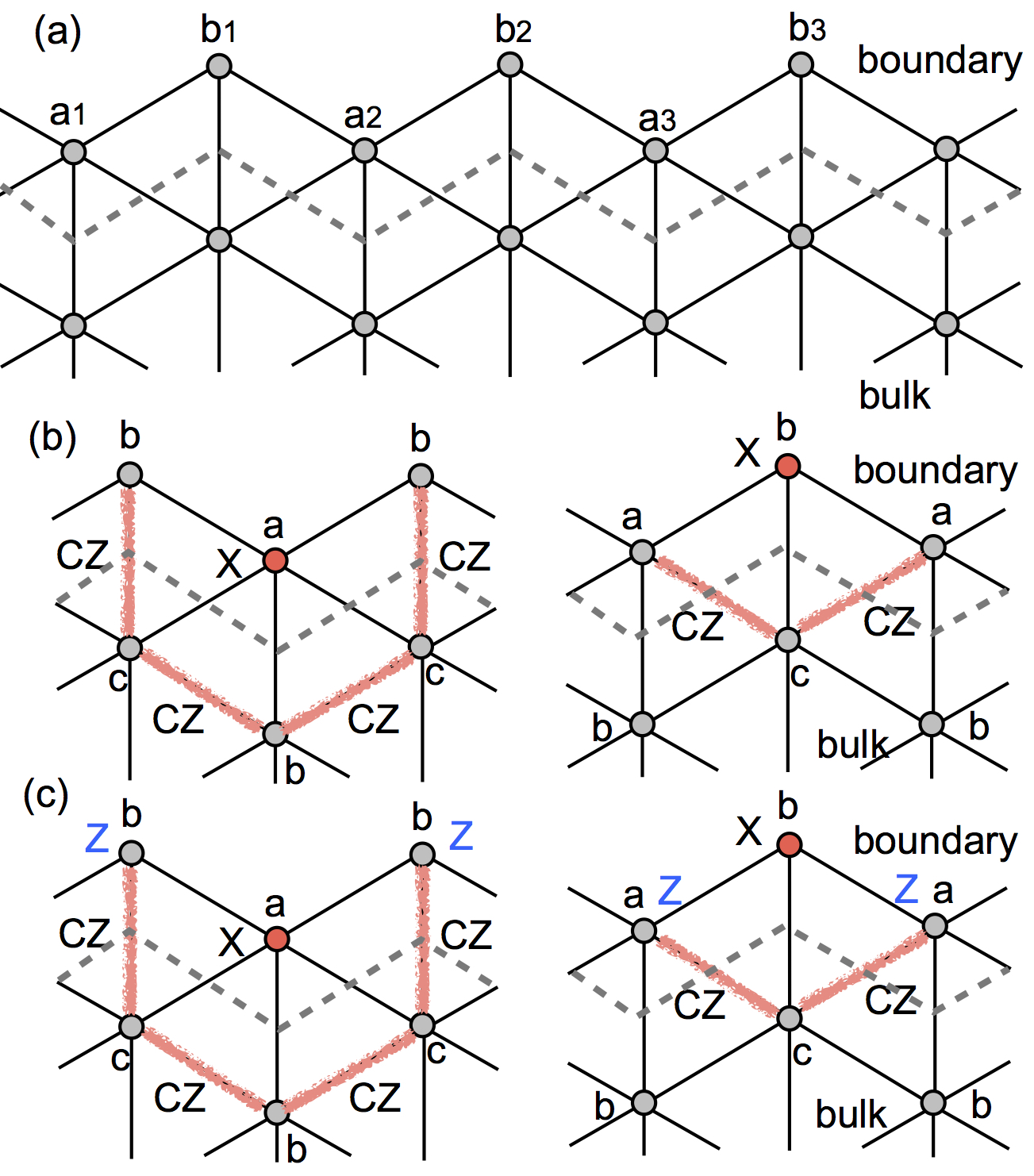}
\caption{(a) Edge and bulk qubits. (b) Dressed boundary operators $\overline{X}_{a}$ and $\overline{X}_{b}$. (c) Dressed boundary operators after conjugation by a symmetry operator $S^{(c)}$. 
} 
\label{fig_2dim_edge}
\end{figure}

We then study the boundary mode for the two-dimensional SPT phases with $\mathbb{Z}_{2}\otimes \mathbb{Z}_{2}\otimes \mathbb{Z}_{2}$ symmetry. Consider the triangular lattice with a boundary as shown in Fig.~\ref{fig_2dim_edge} where \tb{boundary} contains vertices of color $a$ or $b$ on the boundary. We write the Hamiltonian as $H = H_{boundary} + H_{bulk}$ where $H_{bulk}$ consists of all the terms $O_{v}$ with $v \in \tb{bulk}$ and $[H_{boundary},H_{bulk}]=0$. We construct dressed boundary operators which commute with the bulk terms and have proper commutation relations of Pauli operators. Let $U$ be the quantum circuit which is truncated at the boundary such that $U$ is a product of $\mbox{CC}Z$ acting on all the triangles which are dully contained on the lattice. By applying the quantum circuit $U$ to Pauli operators on edge qubits, one can construct dressed boundary operators. Pauli $X$ operators are decorated with control-$Z$ operators involving bulk qubits as shown in Fig.~\ref{fig_2dim_edge}(b) while Pauli $Z$ operators remain unchanged. We shall denote these dressed boundary operators by $\overline{X}_{a_{j}},\overline{X}_{b_{j}},\overline{Z}_{a_{j}},\overline{Z}_{b_{j}}$. 

Let us study how dressed boundary operators transform under symmetry operators, $S^{(a)},S^{(b)},S^{(c)}$:
\begin{equation}
\begin{split}
S^{(a)} &: \overline{X}_{a_{j}} \rightarrow \overline{X}_{a_{j}},  \qquad
\overline{X}_{b_{j}} \rightarrow \overline{X}_{b_{j}},\qquad 
\overline{Z}_{a_{j}}\rightarrow - \overline{Z}_{a_{j}},\qquad
\overline{Z}_{b_{j}}\rightarrow \overline{Z}_{b_{j}}\\
S^{(b)} &: \overline{X}_{a_{j}} \rightarrow \overline{X}_{a_{j}},  \qquad
\overline{X}_{b_{j}} \rightarrow \overline{X}_{b_{j}},\qquad 
\overline{Z}_{a_{j}}\rightarrow \overline{Z}_{a_{j}},\qquad
\overline{Z}_{b_{j}}\rightarrow - \overline{Z}_{b_{j}}\\
S^{(c)} &: \overline{X}_{a_{j}} \rightarrow \overline{Z}_{b_{j-1}} \overline{X}_{a_{j}} \overline{Z}_{b_{j}}, \qquad
\overline{X}_{b_{j}} \rightarrow \overline{Z}_{a_{j}} \overline{X}_{b_{j}} \overline{Z}_{a_{j+1}}, \qquad 
\overline{Z}_{a_{j}}\rightarrow \overline{Z}_{a_{j}},\qquad
\overline{Z}_{b_{j}}\rightarrow \overline{Z}_{b_{j}}.
\end{split}
\end{equation}
From these relations, one finds
\begin{align}
S^{(a)} \sim \prod_{j} \overline{X}_{a_{j}},\qquad
S^{(b)} \sim \prod_{j} \overline{X}_{b_{j}},\qquad
S^{(c)} \sim \prod_{e\in E} \overline{\mbox{C}Z_{e}}
\end{align}
where $E$ represents the set of all the edges on the boundary and the products for $S^{(a)},S^{(b)}$ run over all the vertices of color $a,b$ on the boundary. Let us write down possible boundary terms which respect the symmetry:
\begin{align}
\overline{Z}_{a_{j}}\overline{Z}_{a_{j+1}}, \qquad \overline{Z}_{b_{j}}\overline{Z}_{b_{j+1}},\qquad \overline{X}_{a_{j}} + \overline{Z}_{b_{j-1}} \overline{X}_{a_{j}} \overline{Z}_{b_{j}}, \qquad \overline{X}_{b_{j}} + \overline{Z}_{a_{j}} \overline{X}_{b_{j}} \overline{Z}_{a_{j+1}}.
\end{align} 
The first two terms are ferromagnetic interactions among qubits of color $a$ or $b$ respectively. The third and the last terms lead to the following Hamiltonian at quantum criticality:
\begin{align}
H=-\sum_{j=1}^{2n} \overline{Z}_{j-1}\overline{X}_{j}\overline{Z}_{j+1} - \sum_{j=1}^{2n} \overline{X}_{j}.
\end{align}
This Hamiltonian can be transformed into two decoupled copies of critical quantum Ising model by a duality transformation~\footnote{Consider the following transformation:
$Z_{j-1}X_{j}Z_{j+1}\rightarrow Z_{j-1}Z_{j+1}$ and $X_{j}\rightarrow X_{j}$. Strictly speaking, this duality transformation is well defined only for an infinite system.}. Therefore, one can conclude that the boundary mode may support $\mathbb{Z}_{2}$ ferromagnets (\emph{i.e.} spontaneous breaking of $\mathbb{Z}_{2}\otimes \mathbb{Z}_{2}$), or, $\mathbb{Z}_{2}$ critical models with gapless modes. It is interesting to observe that the above critical Hamiltonian can be written as a sum of one-dimensional SPT Hamiltonians with $\mathbb{Z}_{2}\otimes \mathbb{Z}_{2}$ symmetry: $H=H_{0} + H_{1}$ where $H_{0}= -\sum_{j=1}^{2n} X_{j}$ and $H_{1}=-\sum_{j=1}^{2n} Z_{j-1}X_{j}Z_{j+1}$ and the symmetry operator $S^{(c)}$ is identical to the quantum circuit for a non-trivial $\mathbb{Z}_{2}\otimes \mathbb{Z}_{2}$ SPT phase: $U^{(0,0)}=\prod_{e\in E} \mbox{C}Z_{e}$. Namely, one has $H = H_{0} + U^{(0,0)}H_{0}{U^{(0,0)}}^{\dagger}$.

\subsection{Coupling to gauge fields}

In this subsection and the next, we study the procedure of ``gauging''~\cite{Levin12}. Physically, gauging is a process of minimally coupling a system with global symmetry $G$ to gauge fields with gauge symmetry $G$. Formally, gauging can be viewed as an isometric bijective map (\emph{i.e.} a duality map) from wavefunctions with global symmetry to wavefunctions with gauge symmetry~\cite{Levin12}. Detailed discussions on physical properties of gauged models are beyond the scope of the present paper. Instead, we will present a generic framework of gauging non-trivial models of bosonic SPT phases. We shall see that the use of colorable graphs is particularly convenient in constructing gauged models. 

In this subsection, we briefly review the procedure of $\mathbb{Z}_{2}$ gauging. Consider a square lattice on a torus where qubits live on vertices, and consider the following trivial system with $\mathbb{Z}_{2}$ symmetry:
\begin{align}
H = - \sum_{v} X_{v} \label{eq:trivial}
\end{align}
where $v$ represents vertices. Clearly, this Hamiltonian and its ground state is symmetric under $\mathbb{Z}_{2}$ global transformation $S = \bigotimes_{v} X_{v}$. By applying the $\mathbb{Z}_{2}$ gauging map, one can transform this trivial system with global $\mathbb{Z}_{2}$ symmetry into a two-dimensional system with $\mathbb{Z}_{2}$ gauge symmetry, namely the two-dimensional toric code, as shown below. 

\begin{figure}[htb!]
\centering
\includegraphics[width=0.45\linewidth]{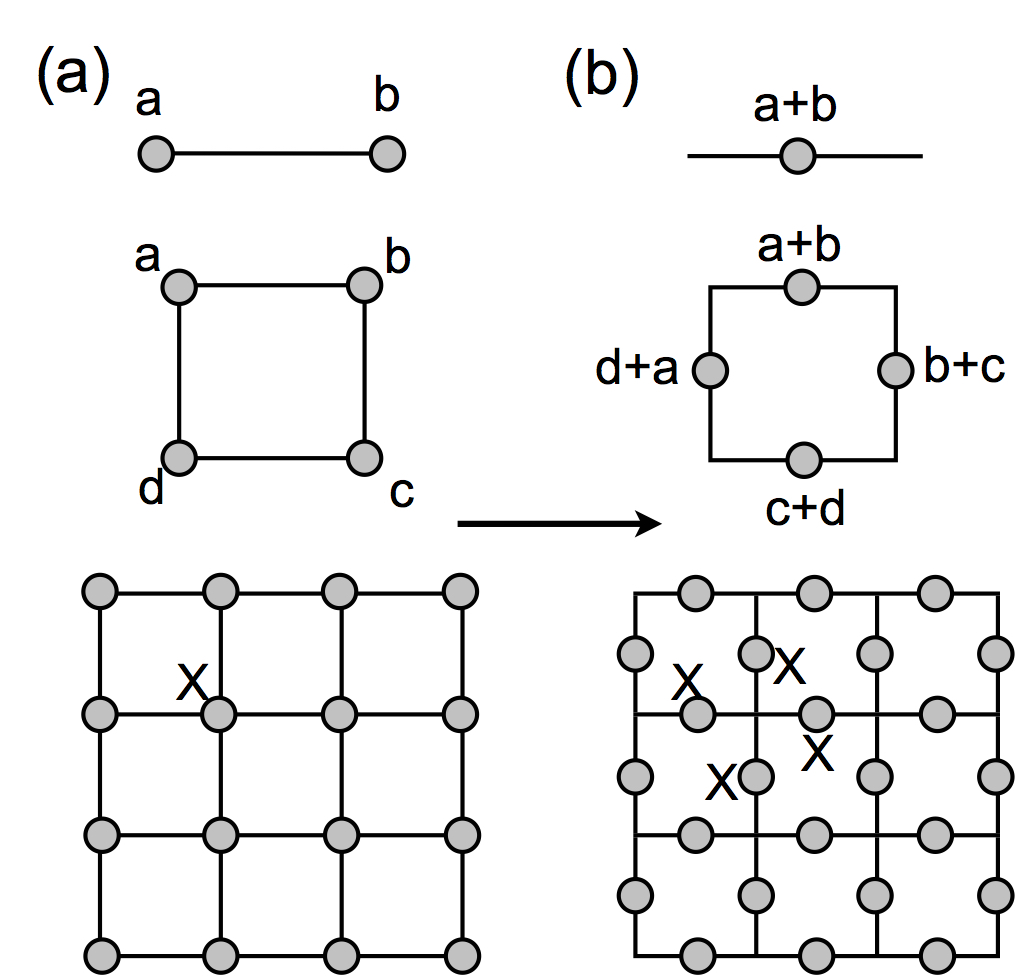}
\caption{Qubits and interaction terms before and after the $\mathbb{Z}_{2}$ gauging. 
} 
\label{fig_gauging}
\end{figure}

To begin, consider a system of two qubits $|a,b\rangle$ where qubits live on endpoints (vertices) of an edge and $a,b=0,1\in\mathbb{Z}_{2}$ (Fig.~\ref{fig_gauging}). Imagine a transformation $(\mathbb{C}^{2})^{\otimes 2}\rightarrow \mathbb{C}^{2}$ whose output is a one-qubit state $|a+b\rangle$ where the summation is modulo $2$ and the output qubit lives on the edge (Fig.~\ref{fig_gauging}). Next, consider a system of qubits supported on vertices of a square lattice and denote the entire Hilbert space by $\mathcal{H}_{0}$. Consider a computational basis state, $|g_{1},g_{2},\ldots,g_{n}\rangle$ where $g_{j}=0,1$, in $\mathcal{H}_{0}$. We think of applying the above transformation to every edge of the system and obtaining an output wavefunction living on edges of the square lattice. This defines a map $\Gamma$ from computational basis states in $\mathcal{H}_{0}$ to computational basis states in $\mathcal{H}_{1}$ where $\mathcal{H}_{1}$ represents the Hilbert space for a system with qubits living on edges of a square lattice. We have
\begin{align}
\Gamma(|g_{1},\ldots,g_{n}\rangle) = |h_{1},\ldots,h_{n'}\rangle, \qquad g_{j},h_{j}=0,1
\end{align}
where $n,n'$ are the numbers of vertices and edges respectively. An important property of $\Gamma$ is that the output wavefunction satisfies the gauge constraints. Namely, for an arbitrary plaquette operator $B_{p}=\prod_{e \in p} Z_{e}$, one has $B_{p}|h_{1},\ldots,h_{n'}\rangle=|h_{1},\ldots,h_{n'}\rangle$.

One can extend the gauging map $\Gamma$ to wavefunctions which are not computational basis states. For this purpose, we will consider some subspaces of $\mathcal{H}_{0}$ and $\mathcal{H}_{1}$. Define the symmetric subspace of $\mathcal{H}_{0}$ as 
\begin{align}
\mathcal{H}_{0}^{sym} = \{ |\psi\rangle \in \mathcal{H}_{0} : S|\psi\rangle = |\psi\rangle   \}
\end{align}
where $S=\bigotimes_{v}X_{v}$. Define the gauge symmetric subspace of $\mathcal{H}_{1}$ as 
\begin{align}
\mathcal{H}_{1}^{sym} = \{ |\psi\rangle \in \mathcal{H}_{1} : B(\gamma)|\psi\rangle = |\psi\rangle \quad \forall \gamma  \}
\end{align}
where $\gamma$ represents an arbitrary closed loop on a square lattice and $B(\gamma)=\prod_{j\in \gamma} Z_{j}$ is a Wilson loop operator. Here we consider not only contractible loops, but also arbitrary closed loops which may have non-trivial winding. We then define the gauging map as follows
\begin{align}
\Gamma(|\psi\rangle) = \frac{1}{\sqrt{2}}\sum_{g_{1},\ldots,g_{n}}C_{g_{1},\ldots,g_{n}}\Gamma( |g_{1},\ldots,g_{n}\rangle )
\end{align}
where $|\psi\rangle=\sum_{g_{1},\ldots,g_{n}}C_{g_{1},\ldots,g_{n}} |g_{1},\ldots,g_{n}\rangle\in \mathcal{H}_{0}^{sym}$. Then one has
\begin{align}
\dim \mathcal{H}_{0}^{sym}= \dim \mathcal{H}_{1}^{sym}.
\end{align}
Even more, the gauging map $\Gamma$ is bijective and isometric (\emph{i.e.} is a duality map). By an isometric map, we mean that the inner product of any pair of wavefunctions is preserved. 

Let us apply the gauging map to a trivial wavefunction $|\psi\rangle$ of Eq.~(\ref{eq:trivial}). Let $|\hat{\psi}\rangle$ be the output wavefunction. We shall see that $|\hat{\psi}\rangle$ is a ground state of the two-dimensional toric code: $\hat{H} = -\sum_{v} A_{v} - \sum_{p} B_{p}$. Indeed, $\mathbb{Z}_{2}$ gauge constraints account for the plaquette terms $B_{p}=\prod_{e\in p} Z_{e}$ in the toric code. As for the star terms $A_{v}=\prod_{v\in e} X_{e}$, observe that $X_{v}|\psi\rangle=|\psi\rangle$ inside $\mathcal{H}_{0}^{sym}$. Flipping a spin at the vertex $v$ is equivalent to flipping four spins on edges that are connected to the vertex $v$ in the gauge theory. Thus, $X_{v}$ operator in $\mathcal{H}_{0}^{sym}$ is equivalent to the vertex term $A_{v}$ in $\mathcal{H}_{1}^{sym}$. As such, the output wavefunction must satisfy $A_{v}|\hat{\psi}\rangle=|\hat{\psi}\rangle$ for all $v$, and thus is a ground state of the toric code. The above procedure can be extended to a $d$-dimensional system with on-site symmetry group $G$ where $G$ is an arbitrary finite group, and the gauging map outputs the $d$-dimensional quantum double model with $G$~\cite{Hu13, Wan15}. 

\subsection{Gauged model and gapped domain wall}

\begin{figure}[htb!]
\centering
\includegraphics[width=0.75\linewidth]{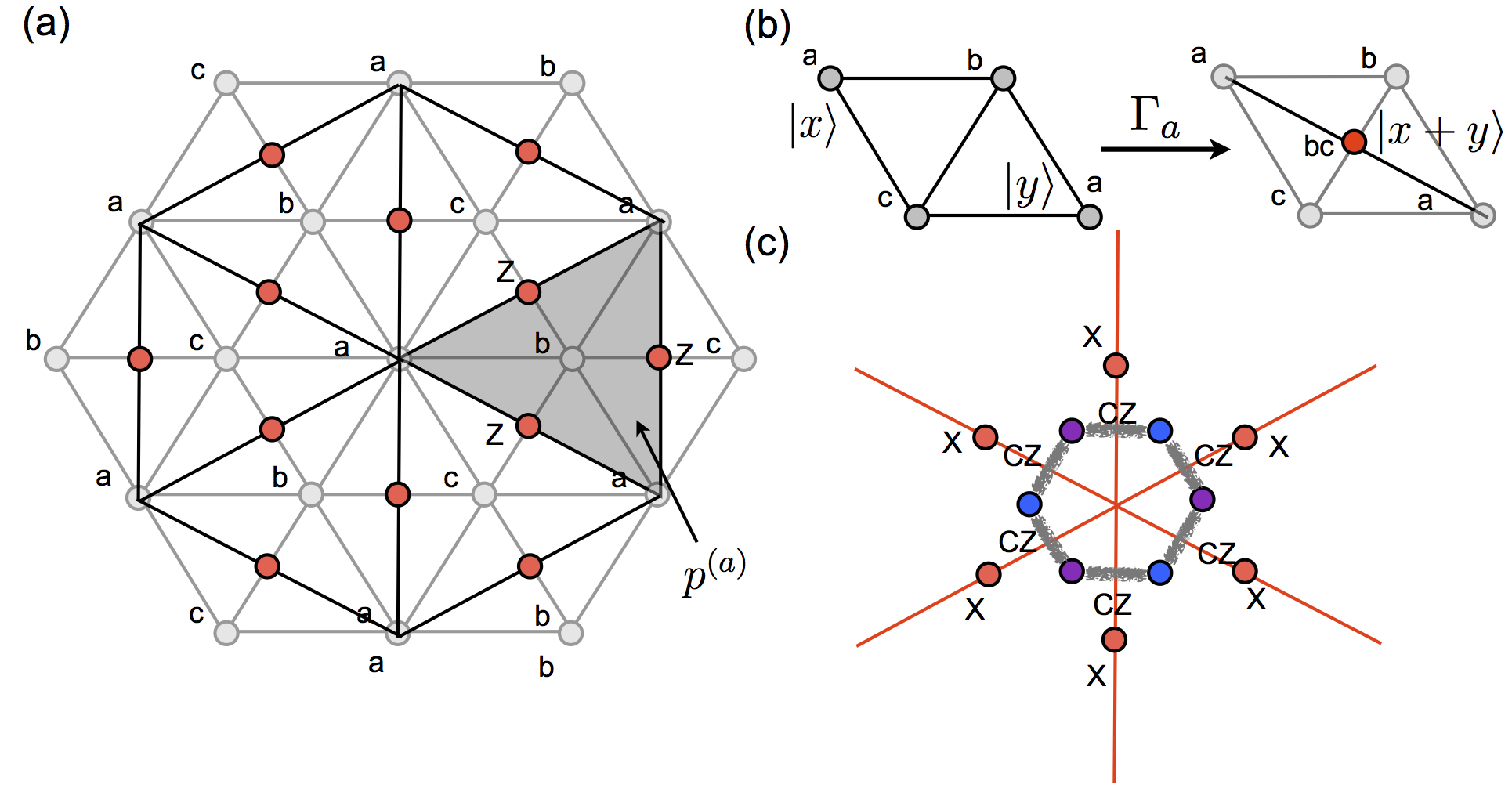}
\caption{(a) Qubits on edges of color $bc$. (b) Gauging maps $\Gamma_{a}$. (c) Interaction terms. 
} 
\label{fig_2dim_gauge}
\end{figure}

In this subsection, we apply the gauging map defined in the previous subsection to SPT wavefunctions. Consider the two-dimensional SPT phase with $\mathbb{Z}_{2}\otimes \mathbb{Z}_{2}\otimes \mathbb{Z}_{2}$ symmetry supported on a three-colorable lattice where qubits are placed on $a,b,c$ vertices respectively. Let us denote the entire Hilbert space by $\mathcal{H}_{0}$. The gauging map can be expressed as follows
\begin{align}
\Gamma = \Gamma_{a} \otimes \Gamma_{b} \otimes  \Gamma_{c}
\end{align}
where $\Gamma_{a}, \Gamma_{b},  \Gamma_{c}$ are $\mathbb{Z}_{2}$ gauging maps acting on qubits living on $a,b,c$ respectively. Recall that edges of a colorable graph can be labeled by pairs of color indices, $ab,bc,ca$, by looking at colors of vertices that are connected by edges. Imagine that we place qubits on $ab,bc,ca$ edges, instead of $a,b,c$ vertices, and denote the entire Hilbert space by $\mathcal{H}_{1}$ (Fig.~\ref{fig_2dim_gauge}(a)). Observe that input wavefunctions of the gauging map $\Gamma_{a}$ live on vertices of color $a$ while its output wavefunctions live on edges of color $bc$ since the middle point of two neighboring vertices of color $a$ is an edge of color $bc$ (Fig.~\ref{fig_2dim_gauge}(b)). Thus, the gauging map $\Gamma$ is a map from computational basis states in $\mathcal{H}_{0}$ to those in $\mathcal{H}_{1}$. A key observation is that, due to the colorability of the graph, one does not need to modify the lattice structure in defining the Hilbert space $\mathcal{H}_{1}$ for the output wavefunctions. One can define $\mathbb{Z}_{2}\otimes\mathbb{Z}_{2}\otimes\mathbb{Z}_{2}$ symmetric subspace $\mathcal{H}_{0}^{sym}$ and $\mathbb{Z}_{2}\otimes\mathbb{Z}_{2}\otimes\mathbb{Z}_{2}$ gauge symmetric subspace $\mathcal{H}_{1}^{sym}$ as before. Then the gauging map $\Gamma=\Gamma_{a}\otimes \Gamma_{b}\otimes \Gamma_{c}$ is a duality map between $\mathcal{H}_{0}^{sym}$ and $\mathcal{H}_{1}^{sym}$.

Let us apply the gauging map $\Gamma$ to the two-dimensional SPT phase. Let $|\psi\rangle$ be a ground state of the two-dimensional SPT Hamiltonian with $\mathbb{Z}_{2}\otimes \mathbb{Z}_{2}\otimes \mathbb{Z}_{2}$ symmetry and $|\hat{\psi}\rangle = \Gamma(|\psi\rangle)$ be the output state. Let $p^{(a)}$ be a plaquette surrounded by vertices of color $a$, as shown in Fig.~\ref{fig_2dim_gauge}(a), and $P^{(a)}$ be the set of such plaquettes. Then one has $B_{p^{(a)}}|\phi\rangle=|\phi\rangle$ where $B_{p^{(a)}}$ is a tensor product of Pauli $Z$ operators surrounding the plaquette $p^{(a)}$. Similar argument holds for $B_{p^{(b)}}$ and $B_{p^{(c)}}$. The original symmetric wavefunction $|\psi\rangle$ satisfies $X_{v^{(a)}}|\psi\rangle=X_{v^{(b)}}|\psi\rangle=X_{v^{(c)}}|\psi\rangle=|\psi\rangle$. After gauging, one can write down the corresponding vertex terms:
\begin{align}
A_{v^{(a)}}|\hat{\psi}\rangle=A_{v^{(b)}}|\hat{\psi}\rangle=A_{v^{(c)}}|\hat{\psi}\rangle=|\hat{\psi}\rangle.
\end{align}
which resemble ordinary vertex terms for the toric code, but are additionally decorated by $\mbox{C}Z$ operators as shown in Fig.~\ref{fig_2dim_gauge}(c). As such, the gauged model can be written as
\begin{align}
\hat{H} =  - \sum_{p\in P^{(a)},P^{(b)},P^{(c)}} B_{p}-  \sum_{v^{(a)}\in V^{(a)}} A_{v^{(a)}}- \sum_{v^{(b)}\in V^{(b)}} A_{v^{(b)}}- \sum_{v^{(c)}\in V^{(c)}} A_{v^{(c)}}.
\end{align}
The Hamiltonian can be viewed as three copies of the toric code which are intricately coupled with each other via $\mbox{C}Z$ phase operators. 

\begin{figure}[htb!]
\centering
\includegraphics[width=0.80\linewidth]{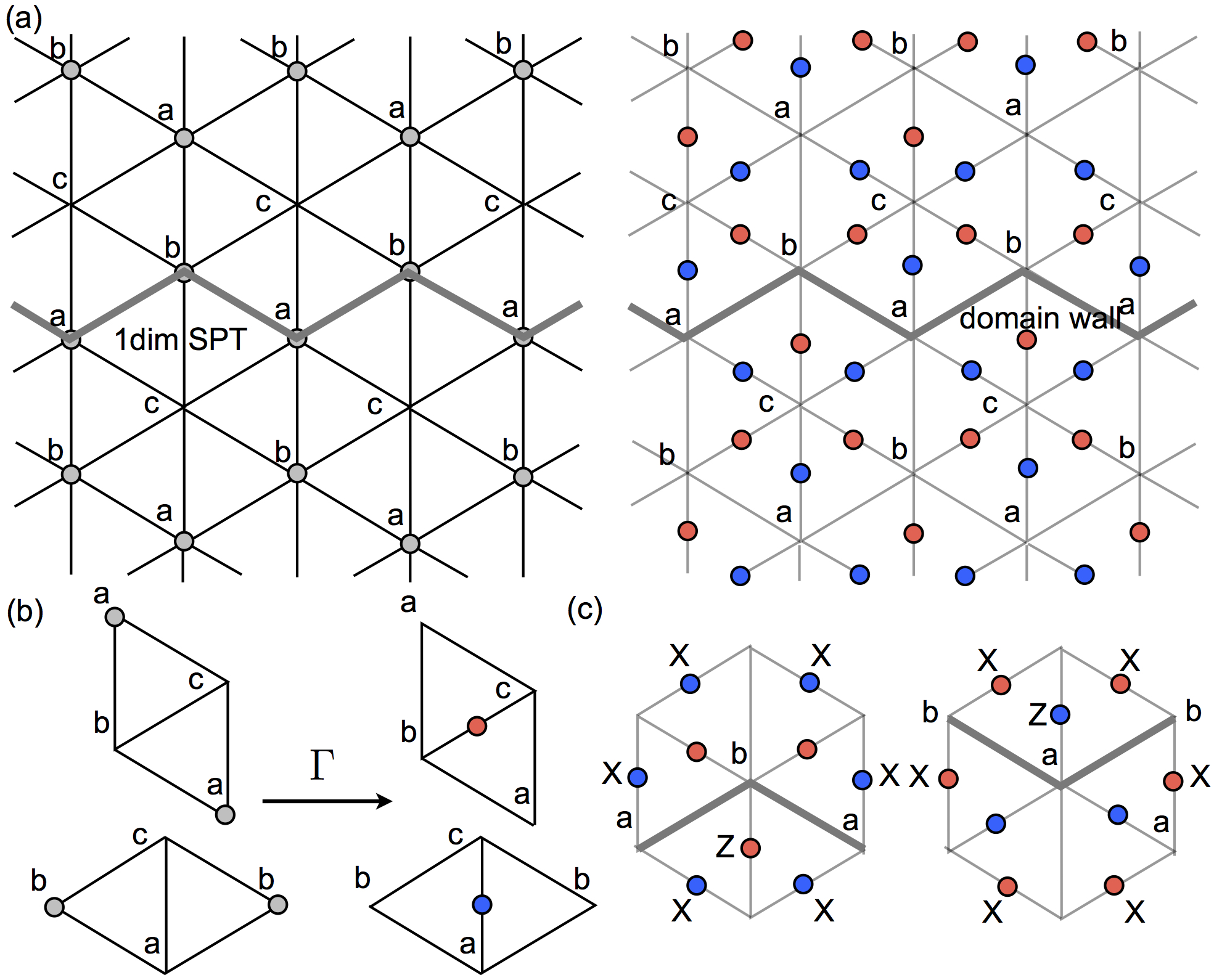}
\caption{(a) A one-dimensional SPT wavefunction in two dimensions and the corresponding gapped domain wall. (b) The gauging map. (c) Terms near the domain wall. 
} 
\label{fig_2dim_domain}
\end{figure}

An interesting application of the gauging map is that one can construct a gapped domain wall in a $d$-dimensional topologically ordered system by using bosonic $(d-1)$-dimensional SPT phases~\cite{Beni15, Beni15c}. To be specific, consider a two-dimensional colorable graph $\Lambda$ with color labels $a,b,c$ where qubits live only on vertices of color $a,b$. Consider a one-dimensional line $\partial \Lambda$ in the graph which consist only of vertices of color $a,b$ as shown in Fig.~\ref{fig_2dim_domain}(a), which splits the entire system into the upper and lower parts. Imagine that a one-dimensional SPT wavefunction with $\mathbb{Z}_{2}\otimes\mathbb{Z}_{2}$ symmetry lives on $\partial \Lambda$ while all the other qubits are in the trivial product state of $|+\rangle$. 

We couple the entire system to $\mathbb{Z}_{2}\otimes \mathbb{Z}_{2}$ gauge fields living in two dimensions. In a gauge theory, qubits live on $bc$-edges and $ca$-edges where the gauging map $\Gamma$ acts on vertices of color $a,b$, and outputs quantum states on $bc$-edges and $ca$-edges as shown in Fig.~\ref{fig_2dim_gauge}(b). The Hamiltonian for the gauged model can be written as
\begin{align}
H= H_{up} + H_{down} + H_{\partial \Lambda}
\end{align}
where $H_{up}, H_{down}$ represent terms on $\Lambda \setminus \partial \Lambda$ while $H_{\partial \Lambda}$ represents terms connecting two Hamiltonians $H_{up}$ and $H_{down}$. Note that $H_{up}, H_{down}$ are identical to those of two copies of the toric code while $H_{\partial \Lambda}$ is different from those of the ordinary toric code and can be viewed as a domain wall connecting $H_{up}$ and $H_{down}$. Terms in $H_{\partial \Lambda}$ are explicitly written in Fig.~\ref{fig_2dim_domain}(c). Let $e_{a},e_{b}$ be electric charges associated with violations of vertex-like Pauli-$X$ terms. Let $m_{a},m_{b}$ be magnetic fluxes associated with violations of plaquette-like Pauli-$Z$ terms. Then labels of anyonic excitations get transformed upon crossing the domain wall as follows:
\begin{align}
m_{a} \rightarrow m_{a}e_{b} ,\qquad m_{b} \rightarrow m_{b}e_{a} ,\qquad 
e_{a} \rightarrow e_{a} ,\qquad e_{b} \rightarrow e_{b}. 
\end{align}

The fact that the domain wall transposes labels of anyonic excitations implies that the one-dimensional SPT wavefunction is non-trivial and cannot be created from a trivial state by a symmetric finite-depth quantum circuit~\cite{Beni15}. The key observation is that, while the domain wall is localized along a one-dimensional region, it cannot be created by a local unitary transformation acting on qubits in the neighborhood of the domain wall. Suppose there exists a local unitary $U$ which creates the domain wall by acting only on qubits in the neighborbood of the domain wall. Let $\ell$ be a string operator corresponding to the propagation of a magnetic flux in the absence of the domain wall. Then $U\ell U^{\dagger}$ differs from $\ell$ only at the intersection with the domain wall. This implies that a magnetic flux remains to be a magnetic flux upon crossing the domain wall, leading to a contradiction. Thus, to create the domain wall, one needs to apply a local unitary transformation on all the qubits on one side of the system. This argument enables us to show that the underlying one-dimensional SPT wavefunction is non-trivial. Suppose that there exists a symmetric local unitary transformation $\tilde{U}$ which creates the underlying SPT wavefunction. After gauging this symmetric unitary operator, one obtains local unitary transformation $U$ defined in the gauge theory which is localized along the domain wall, leading to a contradiction. In the above argument, braiding statistics of anyons plays the role of topological invariants in proving the non-triviality of SPT wavefunctions. 

Finally, we briefly comment on a gapped domain wall which can be constructed by gauging the two-dimensional SPT wavefunction in three dimensions. The domain wall connects three copies of the three-dimensional toric code where electric charges $e_{1},e_{2},e_{3}$ are point-like while magnetic fluxes $m_{1},m_{2},m_{3}$ are loop-like. Upon crossing the domain wall, electric charges remain unchanged:
\begin{align}
e_{1}\rightarrow e_{1} \qquad e_{2}\rightarrow e_{2} \qquad e_{3}\rightarrow e_{3}
\end{align}
while magnetic fluxes transform into composites of magnetic-fluxes and loop-like superpositions of electric charges:
\begin{align}
m_{1}\rightarrow m_{1}s_{23} \qquad m_{2}\rightarrow m_{2}s_{13} \qquad m_{3}\rightarrow m_{3}s_{12}. 
\end{align}
Here $s_{ij}$ ($i\not=j$) is a one-dimensional excitation which is a superposition of electric charges $e_{i}$ and $e_{j}$. It has been found that $s_{ij}$ can be characterized by a non-trivial wavefunction of a one-dimensional $\mathbb{Z}_{2}\otimes \mathbb{Z}_{2}$ bosonic SPT phase~\cite{Beni15}. Namely, if the emerging wavefunction is written as a superposition of excited eigenstates, its expression is identical to a fixed-point wavefunction of a one-dimensional non-trivial SPT phase. It is important to note that these fluctuating charges $s_{ij}$ are loop-like objects which are unbreakable in a sense that their creation requires membrane-like operators. In addition, $m_{i}$ and $s_{jk}$ exhibit non-trivial three-loop braiding statistics~\footnote{In a three-loop braiding, two loops are braided while the third loop pierces through two loops~\cite{Wang14}.}. 

\subsection{Higher-dimensional generalization}

Finally, we briefly present construction of $d$-dimensional SPT phases with ${\mathbb{Z}_{2}}^{\otimes d+1}$ symmetry. Consider a $d$-dimensional simplicial lattice which is $(d+1)$-colorable with color labels $a_{1},\ldots,a_{d+1}$ and place qubits on each vertex. Let $H_{0}^{(0,0,\ldots)}=-\sum_{v\in V}X_{v}$ be a trivial Hamiltonian. The non-trivial Hamiltonian $H_{1}$ can be constructed from the following finite-depth quantum circuit:
\begin{align}
H^{(0,0,\ldots)}_{1}=U^{(0,0,\ldots)}H^{(0,0,\ldots)}_{0}{U^{(0,0,\ldots)}}^{\dagger},\qquad U^{(0,0,\ldots)} = \prod_{(i_{1},i_{2},\ldots,i_{d+1})\in \Delta} \mbox{C}^{\otimes d} Z_{i_{1},i_{2},\ldots,i_{d+1}}
\end{align}
where $\Delta$ represents the set of all the $d$-simplexes. The interaction terms of $H_{1}^{(0,0,\ldots)}$ are 
\begin{align}
H_{1}^{(0,0,\ldots)}= - \sum_{v\in V} O_{v},\qquad O_{v}= X_{v}\prod_{(i_{1},i_{2},\ldots,i_{d})\in (d-1)\tb{-link}(v)} \mbox{C}^{\otimes d-1} Z_{i_{1},i_{2},\ldots ,i_{d}}
\end{align}
where $(d-1)\tb{-link}(v)$ represents the set of all the $(d-1)$-links of the vertex $v$. Namely, a $(d-1)$-link of $v$ is a $(d-1)$-simplex which forms a $d$-complex by adding the vertex $v$. There are $d+1$ copies of $\mathbb{Z}_{2}$ symmetry operators associated with color labels $a_{j}$:
\begin{align}
S^{(j)} = \prod_{v^{(a_{j})}\in V^{(a_{j})} } X_{v^{(a_{j})}} = \prod_{v^{(a_{j})}\in V^{(a_{j})}} O_{v^{(a_{j})}}\qquad j=1,\ldots,d+1
\end{align}
where $V^{(a_{j})}$ represents the set of all the vertices of color $a_{j}$. 

Boundary mode in higher-dimensional SPT phases can be studied in a similar manner. Namely, the $(d-1)$-dimensional boundary can support a quantum critical Hamiltonian of the form $H = H^{(0,0,\ldots)}_{0} + H^{(0,0,\ldots)}_{1}$ where $H^{(0,0,\ldots)}_{0}$ and $H^{(0,0,\ldots)}_{1}$ are trivial and non-trivial $d-1$-dimensional SPT Hamiltonians with $\mathbb{Z}_{2}^{\otimes d}$ symmetry respectively. The gauged model can be defined on the same colorable lattice where qubits are placed on centers of $(d-1)$-simplexes instead of vertices. The model looks like $d+1$ copies of the toric code whose vertex terms have decorations of $\mbox{C}^{\otimes d-1}Z$ operators, mixing $d+1$ copies in an intricate way. If one gauges the $d$-dimensional SPT phase in $d+1$ dimensions, one obtains a gapped domain wall in $d+1$ copies of the toric code where codimension-$1$ magnetic flux get transformed into a composite of codimension-$1$ magnetic flux and codimension-$1$ fluctuating superpositions of electric charge upon crossing the domain wall~\cite{Beni15}.

\section{Topological phases with generalized global symmetries}\label{sec:higher-form}

In this section, we present examples of bosonic SPT phases with higher-form global symmetries. The key distinction between models with $0$-form symmetries and higher-form symmetries is that qubits are placed on $q$-simplexes for models with $q$-form symmetries. We use the multi-qubit control-$Z$ gate to construct non-trivial SPT wavefunctions with $q$-form symmetries. We also study the boundary mode, gauged models and corresponding gapped domain walls. 

\subsection{Three-dimensional model with $1$-form symmetries}

\begin{figure}[htb!]
\centering
\includegraphics[width=0.5\linewidth]{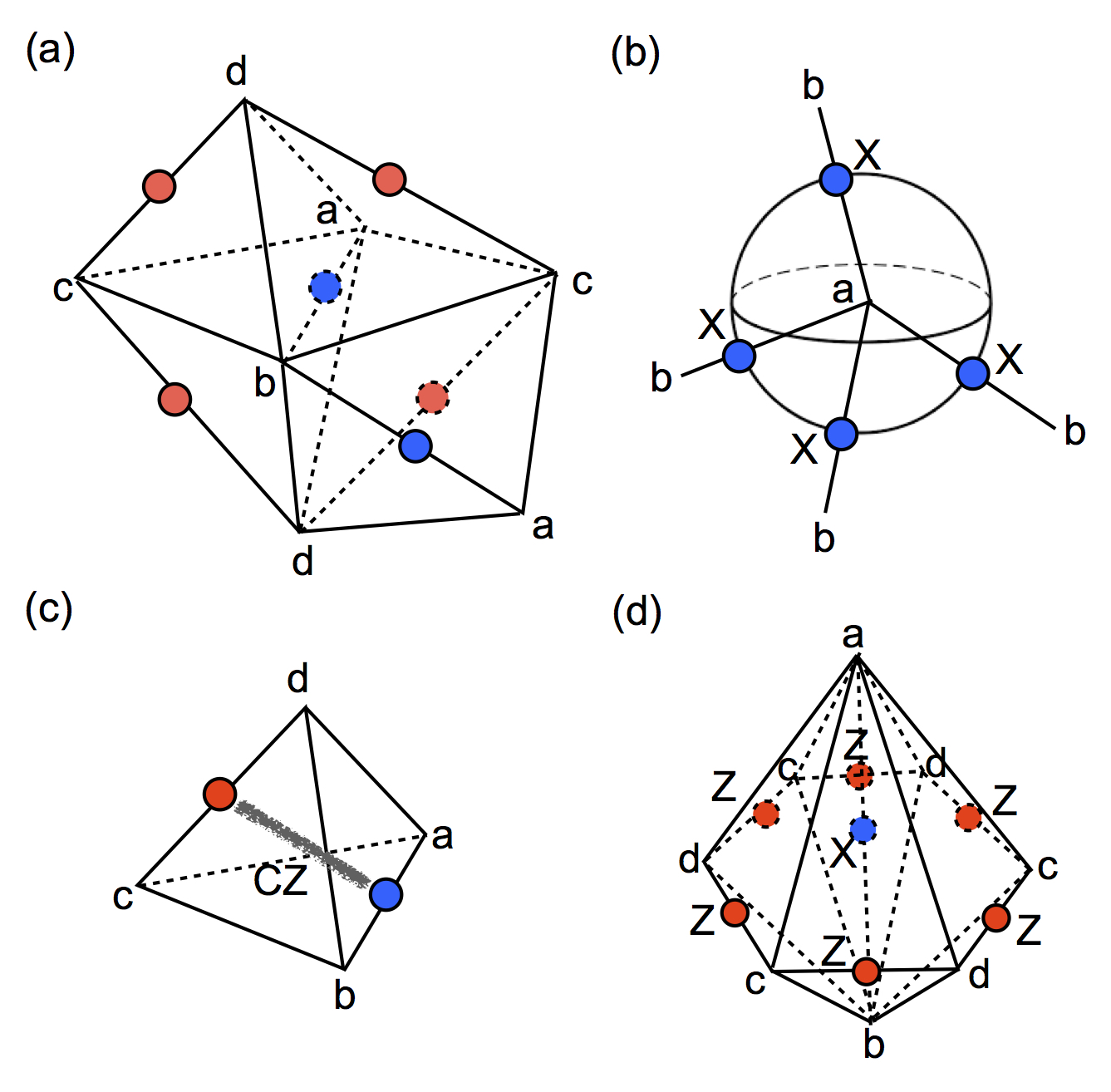}
\caption{Three-dimensional SPT phase with $1$-form $\mathbb{Z}_{2}\otimes \mathbb{Z}_{2}$ symmetry. (a) Qubits on a colorable graph. (b) A symmetry operator associated with a closed sphere. (c) A finite-depth quantum circuit. (d) Interaction terms. A Pauli-$X$ is decorated by Pauli-$Z$ operators acting on $1$-links.
} 
\label{fig_3dim_1form}
\end{figure}

In this subsection, we present a model of a three-dimensional SPT phase with $1$-form $\mathbb{Z}_{2}\otimes \mathbb{Z}_{2}$ symmetry. Consider a three-dimensional simplicial lattice $\Lambda$ which is four-colorable with color labels $a,b,c,d$. Edges of the graph are labeled by pairs of colors, such as $ab,ac,ad,\ldots$. We place qubits on $ab$-edges and $cd$-edges of $\Lambda$ while no qubits are placed on edges of other colors or vertices as shown in Fig.~\ref{fig_3dim_1form}(a). 

We specify global symmetry operators which have geometries of two-dimensional closed manifold. Consider a closed $2$-manifold $\mathcal{M}$ which intersects with edges of $\Lambda$. One may view $\mathcal{M}$ as a two-dimensional simplicial sublattice of $\Lambda$. We place $\mathcal{M}$ such that it does not intersect with vertices. (In other words, $\mathcal{M}$ is a discretization of a plane on a dual lattice). Let $S(\mathcal{M})$ be a sheet of Pauli $X$ operators acting on edges which are intersected by $\mathcal{M}$. The simplest example of such a global symmetry operator can be constructed by considering a small $2$-sphere as shown in Fig.~\ref{fig_3dim_1form}(b). There are two different types of symmetry operators, acting on $ab$-edges and $cd$-edges respectively, and they are separable. Namely, one can choose $\mathcal{M}$ such that it does not intersect with $cd$-edges. One can also glue $2$-spheres, which are centered at vertices of color $a$ and $b$, together to construct a sheet which intersects only with $ab$-edges. So the system has $1$-form $\mathbb{Z}_{2}\otimes \mathbb{Z}_{2}$ symmetry. 

We then construct the model Hamiltonian~\footnote{Our model seems very similar to the model proposed in~\cite{Burnell14}, which was treated as a bosonic $0$-form SPT Hamiltonian protected by time-reversal symmetry. }. First the trivial Hamiltonian is given by $H_{0} = - \sum_{e\in E} X_{e}$ where $E$ represents the set of all the edges. The non-trivial Hamiltonian is obtained by applying $\mbox{C}Z$ operators on every pair of qubits inside each $3$-simplex. Let us denote such a unitary operator by
\begin{align}
U^{(1,1)} = \prod_{(i,j)\in \Delta} \mbox{CZ}_{i,j}
\end{align}
where $\Delta$ represents the set of all the $3$-simplexes and $\mbox{CZ}_{i,j}$ acts on two qubits inside a $3$-simplex (Fig.~\ref{fig_3dim_1form}(c)). The non-trivial Hamiltonian $H_{1}=U^{(1,1)}H_{0}{U^{(1,1)}}^{\dagger}$ is given by 
\begin{align}
H_{1} = - \sum_{e\in E} O_{e}, \qquad O_{e}=X_{e} \prod_{e' \in 1\tb{-link}(e)} Z_{e'}
\end{align}
where $1\tb{-link}(e)$ represents the set of $1$-links of an edge $e$ (Fig.~\ref{fig_3dim_1form}(d)). Since $U^{(1,1)}$ is a finite-depth quantum circuit, $H_{0}$ and $H_{1}$ belong to the same topological phase in the absence of symmetry. 
Yet, each local component in $U^{(1,1)}$ does not commute with global symmetry operators. We claim that ground states of $H_{0}$ and $H_{1}$ cannot be connected by a symmetric local unitary.

One can see that interaction terms $O_{e}$ commute with $1$-form symmetry operator $S(\mathcal{M})$ for any closed $2$-manifold $\mathcal{M}$. Let us verify that the ground state $|\psi_{1}\rangle$ of $H_{1}$ is symmetric: $S(\mathcal{M})|\psi_{1}\rangle=|\psi_{1}\rangle$ for all $\mathcal{M}$. We prove this for the cases where $\mathcal{M}$ is a contractible sphere. Let $\mathcal{M}_{v}$ be a small $2$-sphere surrounding the vertex $v$. With some speculation, one can confirm that
\begin{align}
\prod_{v\in e}O_{e} = S(\mathcal{M}_{v})
\end{align}
since all the Pauli $Z$ operators cancel with each other. Since $O_{e}|\psi_{1}\rangle=|\psi_{1}\rangle$, one has $S(\mathcal{M}_{v})|\psi_{1}\rangle=|\psi_{1}\rangle$. Any contractible sphere $\mathcal{M}$ can be constructed by attaching $\mathcal{M}_{v}$ for various $v$, and thus $S(\mathcal{M})|\psi_{1}\rangle=|\psi_{1}\rangle$.

Under $1$-form global symmetry, excitations are loop-like objects. Namely, to create excitations by operators which commute with global symmetry operators, one needs to consider closed strings of Pauli $Z$ operators. Let $\gamma^{ab}$ be a closed loop consisting of $ab$-edges and let $Z(\gamma^{ab})$ be a string of Pauli $Z$ operators acting on $\gamma^{ab}$. This operator creates string-like excitations, violating $O_{e}$ along $\gamma^{ab}$ while commuting with symmetry operators. A similar operator $Z(\gamma^{cd})$ can be defined for a closed-loop $\gamma^{cd}$ of $cd$-edges. The $1$-form symmetry, imposed by $S(\mathcal{M})$, can be viewed as a conservation law for loop-like excitations where the number of the cuts of loop-like excitations made by $\mathcal{M}$ must be even. As such, there are two copies of $\mathbb{Z}_{2}$ conservation law on loop-like excitations supported on $ab$-edges and $cd$-edges. 

\begin{figure}[htb!]
\centering
\includegraphics[width=0.75\linewidth]{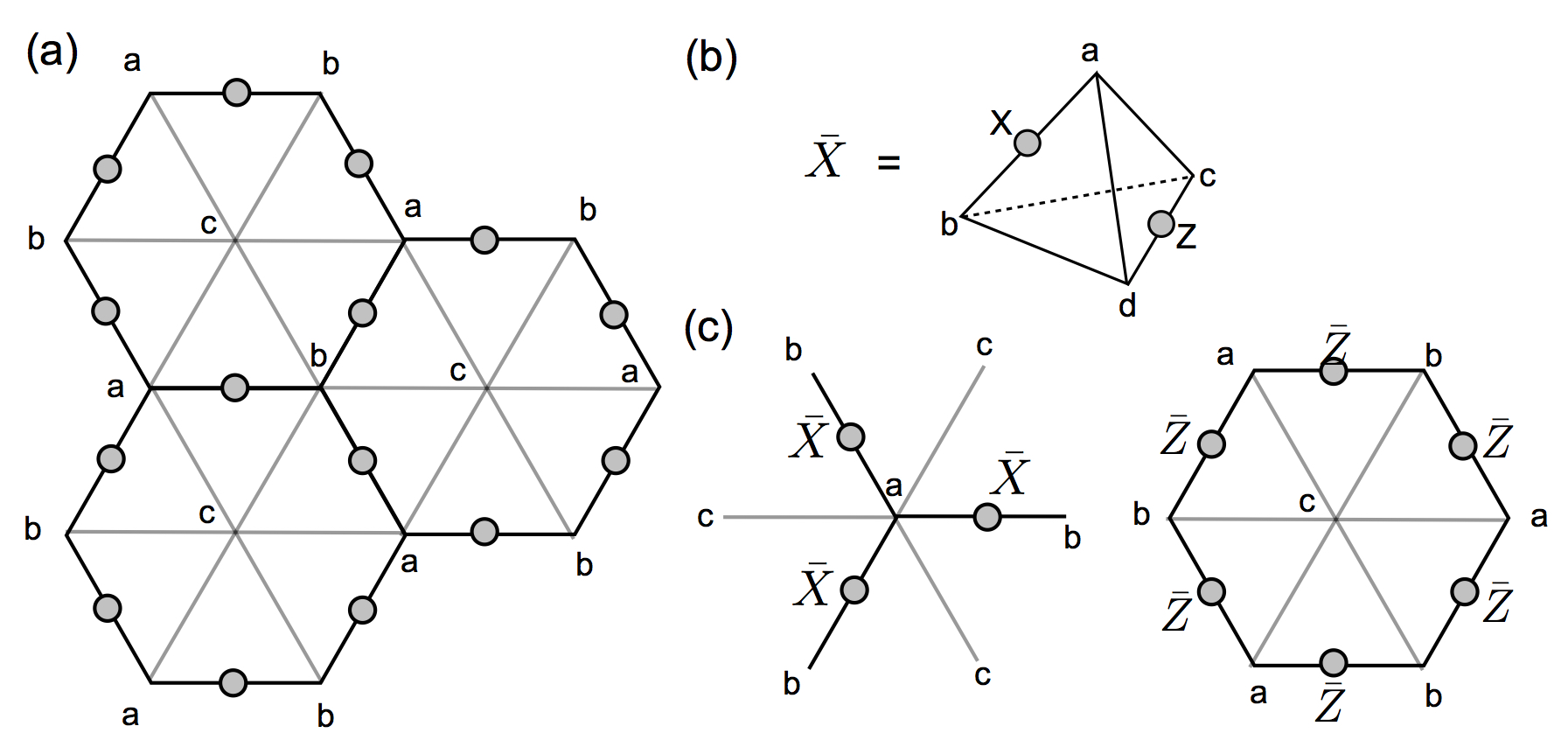}
\caption{(a) Qubits on the boundary. (b) Dressed boundary operator. A vertex of color $d$ is contained inside the bulk while vertices of color $a,b,c$ are on the boundary. (c) Symmetry operators on the boundary. $\bar{X}$ and $\bar{Z}$ indicate that these operators are expressed in terms of dressed boundary operators.
} 
\label{fig_2dim_edge_1form}
\end{figure}

We then study the boundary mode of the aforementioned three-dimensional model. We choose the two-dimensional boundary which consists only of vertices of color $a,b,c$ such that the boundary can be viewed as a three-colorable graph with qubits living on $ab$-edges (Fig.~\ref{fig_2dim_edge_1form}(a)). As before, we are interested in the Hamiltonian with the boundary term of the form: $H = H_{bulk} + H_{boundary}$ where $[H_{bulk},H_{boundary}]=0$, and the low-energy subspace is denoted by $\mathcal{C}$. Dressed boundary operators, denoted by $\overline{X}_{e^{(ab)}},\overline{Z}_{e^{(ab)}}$ for $ab$-edge $e^{(ab)}$, can be found by applying the truncated quantum circuit $U$ to Pauli operators associated with qubits on the boundary. We find that $\overline{X}_{e^{(ab)}}$ is decorated with a Pauli $Z$ operator on the bulk while $\overline{Z}_{e^{(ab)}}=Z_{e^{(ab)}}$ remains unchanged (Fig.~\ref{fig_2dim_edge_1form}(b)). 

Let us study how symmetry operators $S(\Gamma_{ab})$ and $S(\Gamma_{cd})$ act on dressed boundary operators. Let us take $\Gamma_{ab}$ to be a sphere whose center is at the vertex $v$ of color $a$ or $b$ on the boundary. Then $S(\Gamma_{ab})$ acts as a vertex-like term as depicted in Fig.~\ref{fig_2dim_edge_1form}(c). As for $S(\Gamma_{cd})$, consider a sphere with the center being at a vertex of color $c$ on the boundary. Then $S^{(cd)}$ is a plaquette-like term as shown in Fig.~\ref{fig_2dim_edge_1form}(c). Thus the boundary mode supports topologically ordered states, namely the two-dimensional toric code while the bulk is trivial. The non-triviality of the boundary mode is a strong indication of the non-triviality of the three-dimensional SPT wavefunction. We shall present additional supports for the non-triviality below.

\subsection{Gauging $1$-form symmetries}

\begin{figure}[htb!]
\centering
\includegraphics[width=0.55\linewidth]{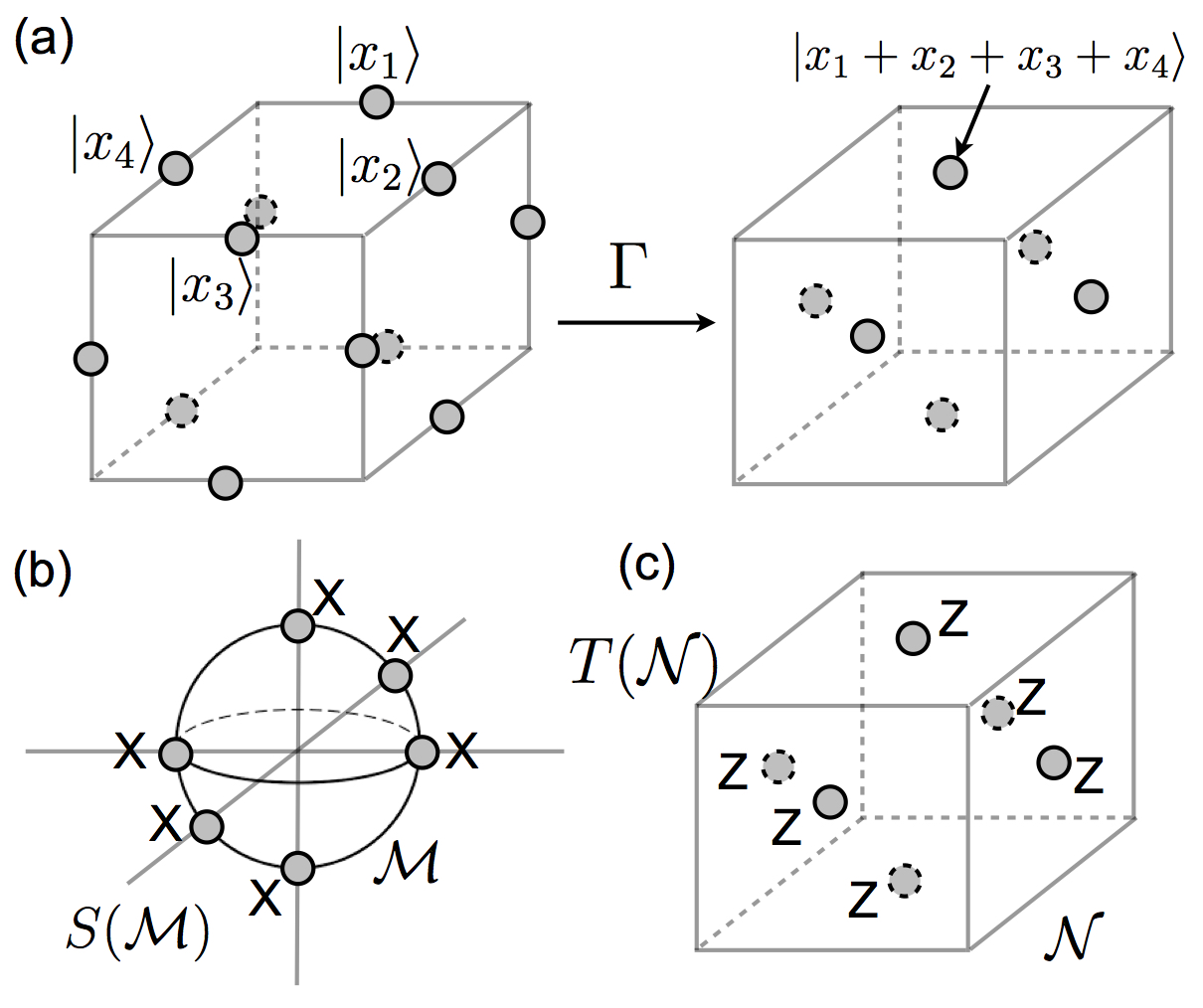}
\caption{(a) Gauging the $1$-form $\mathbb{Z}_{2}$ symmetry. (b) A $1$-form symmetry operator. (c) A generalized gauge constraint. 
} 
\label{fig_1form_gauge}
\end{figure}

We have seen that one can couple $0$-form SPT phases to gauge fields where physical degrees of freedom lives on edges of a graph with gauge constraints acting on plaquettes. In the cases of $1$-form SPT phases, one needs to modify gauge constraints as shown in Fig.~\ref{fig_1form_gauge}(a) where physical degrees of freedom live on plaquettes and gauge constraints act on volumes. In this subsection, we illustrate the procedure of gauging $1$-form symmetries. 

Let us define the gauging map $\Gamma$ for systems with $1$-form $\mathbb{Z}_{2}$ symmetries. To begin, consider a system of $12$ qubits which live on edges of a single cube as shown in Fig.~\ref{fig_1form_gauge}(a). Consider a computational basis state of the form $|x_{1},x_{2},\ldots,x_{12}\rangle$ where $x_{j}=0,1$. The output state of the gauging map $\Gamma$ is a $6$-qubit state where qubits live on plaquettes of the cube. Namely, the spin values of the output state are given by $\mathbb{Z}_{2}$ summation of spin values on edges surrounding the plaquettes as shown in Fig.~\ref{fig_1form_gauge}(a). One can extend the above map to arbitrary three-dimensional lattices. Let $\mathcal{H}_{1}$ and $\mathcal{H}_{2}$ be the Hilbert spaces for the systems where qubits live on edges and plaquettes of the cubic lattice respectively. Then $\Gamma$ can be viewed as a map from computational basis states in $\mathcal{H}_{1}$ to those in $\mathcal{H}_{2}$. A key observation is that the output wavefunctions always satisfy generalized gauge constraints. Namely, if one sums up all the spin values on plaquettes surrounding a cube, then one obtains $0$ modulo $2$, implying $\mathbb{Z}_{2}$ generalized gauge symmetry as in Fig.~\ref{fig_1form_gauge}(c). More generically, let $\mathcal{N}$ be a closed $2$-manifold consisting of plaquettes of the lattice. We define the generalized gauge symmetry operator by $T(\mathcal{N})$ which implements Pauli $Z$ operators on plaquettes contained in $\mathcal{N}$. Then the output wavefunction $|\hat{\psi}\rangle$ always satisfies $T(\mathcal{N})|\hat{\psi}\rangle=|\hat{\psi}\rangle$ for all $\mathcal{N}$.

One can view $\Gamma$ as a duality map by restricting our attentions to some proper subspaces of $\mathcal{H}_{1}$ and $\mathcal{H}_{2}$. Let $\mathcal{H}_{1}^{sym}\subset \mathcal{H}_{1}$ be a subspace of wavefunctions which are symmetric under $1$-form symmetry:
\begin{align}
\mathcal{H}_{1}^{sym} = \{ |\psi\rangle \in \mathcal{H}_{1} : S(\mathcal{M})|\psi\rangle=|\psi\rangle \ \forall \mathcal{M}  \}. 
\end{align}
Here $1$-form symmetry operators can be written as $S(\mathcal{M})$ for an arbitrary closed $2$-manifold $\mathcal{M}$ where $S(\mathcal{M})$ implements Pauli $X$ operators on edges intersected by $\mathcal{M}$ (Fig.~\ref{fig_1form_gauge}(b)). Similarly, we define the gauge symmetric subspace as follows:
\begin{align}
\mathcal{H}_{2}^{sym} = \{ |\hat{\psi}\rangle \in \mathcal{H}_{2} : T(\mathcal{N})|\hat{\psi}\rangle=|\hat{\psi}\rangle  \ \forall \mathcal{N} \}. 
\end{align}
Then one can see that $\Gamma$ induces a duality map between $\mathcal{H}_{1}^{sym}$ and $\mathcal{H}^{sym}_{2}$. Note that closed $2$-manifold $\mathcal{M}, \mathcal{N}$ are defined for the dual and direct lattices respectively. 

Let us apply the $1$-form $\mathbb{Z}_{2}$ gauging map to the trivial three-dimensional Hamiltonian: $H=-\sum_{e}X_{e}$ with a trivial ground state $|\psi\rangle=|+\rangle^{\otimes n}$. The system respects $1$-form $\mathbb{Z}_{2}$ symmetry since the Hamiltonian consists only of Pauli $X$ operators. The original ground state $|\psi\rangle$ satisfies $X_{e}|\psi\rangle=|\psi\rangle$ for all $e$. Upon gauging, the output wavefunction $|\hat{\psi}\rangle$ satisfies $A_{e}|\hat{\psi}\rangle=|\hat{\psi}\rangle$ where $A_{e}$ is a tensor product of Pauli $X$ operators acting on plaquettes attached to an edge $e$. Thus the output wavefunction $|\hat{\psi}\rangle$ is a ground state of the $(2,1)$-toric code. Recall that, if one gauges the trivial three-dimensional Hamiltonian as a model with $0$-form symmetry, one obtains the $(1,2)$-toric code. In higher dimensions, the gauged models for the trivial Hamiltonian with $0$-form and $1$-form symmetry are the $(1,d-1)$-toric code and the $(2,d-2)$-toric code respectively. 

\begin{figure}[htb!]
\centering
\includegraphics[width=0.55\linewidth]{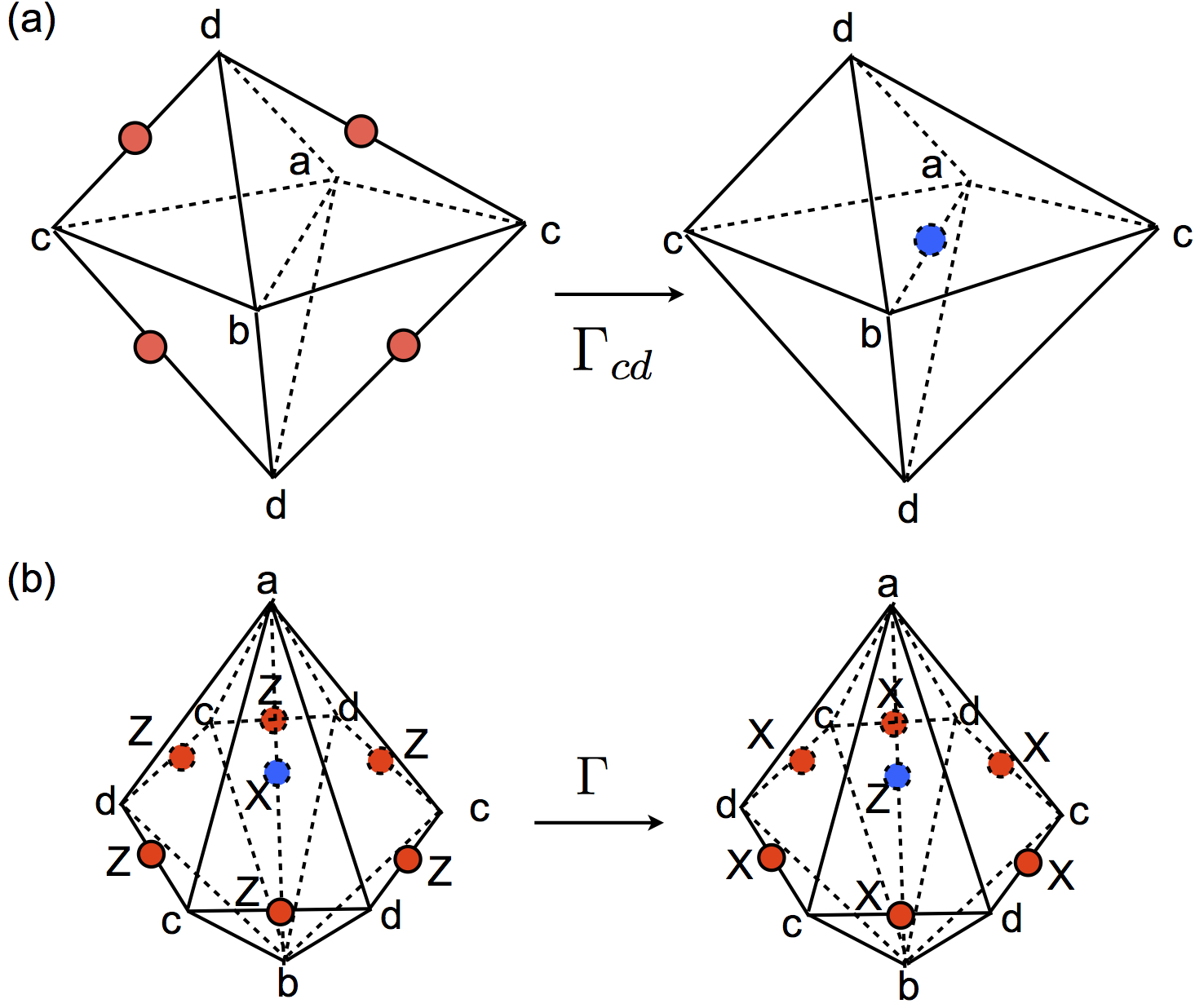}
\caption{(a) A gauging map $\Gamma_{cd}$. (b) $O_{e}$ before and after gauging.  
} 
\label{fig_11_gauge}
\end{figure}

Now, we shall apply the gauging map to the non-trivial three-dimensional Hamiltonian $H_{1}$ with $1$-form symmetry. One can write the gauging map as $\Gamma=\Gamma_{ab}\otimes \Gamma_{cd}$. Consider a plaquette consisting of $ab$-edges as shown in Fig.~\ref{fig_11_gauge}(a). Let $|x_{1},\ldots,x_{2n}\rangle$ be a computational basis state supported on $ab$-edges on the plaquette. The gauging map $\Gamma_{ab}$, when acting on these qubits, outputs a single-qubit state $|x_{1}+x_{2}+\ldots+x_{2n}\rangle$ that lives at the center of the plaquette where summation is modulo $2$. Notice that the center of a plaquette of $cd$-edges coincides with the center of a $ab$-edge as shown in Fig.~\ref{fig_11_gauge}(a). Thus, for qubits supported on $cd$-edges, the output wavefunction is supported on qubits on $ab$-edges, and vice versa. Note that one does not need to modify the lattice structure in order to construct the gauged model. 

Let us find the gauged model of $H_{1}$. The original ground state wavefunction $|\psi_{1}\rangle$ satisfies $O_{e}|\psi_{1}\rangle = |\psi_{1}\rangle$. The corresponding operators $\hat{O}_{e}$ are shown in Fig.~\ref{fig_11_gauge}(b) which are identical to the original operators after exchanging Pauli $X$ and $Z$ operators. Namely, the gauged model can be written as 
\begin{align}
\hat{H}_{1} = U_{H} H_{1} U_{H}^{\dagger}
\end{align}
where $U_{H}$ represents a transversal Hadamard operator. This implies that the gauged model $\hat{H}_{1}$ is not topologically ordered. Since $\hat{H}_{0}$ is topologically ordered while $\hat{H}_{1}$ is not, $H_{0}$ and $H_{1}$ belong to different quantum phases in the presence of $1$-form symmetry. 

Next, we shall gauge the non-trivial three-dimensional Hamiltonian $H_{1}$ in four dimensions and construct a gapped domain wall for the four-dimensional toric code. Consider a four-dimensional five-colorable graph with color labels $a,b,c,d,e$ where qubits are placed on $ab$-edges and $cd$-edges. Choose a codimension-$1$ hypersurface (a three-dimensional volume) $\partial \Lambda$ and place the non-trivial $1$-form SPT wavefunction with $\mathbb{Z}_{2}\otimes \mathbb{Z}_{2}$ symmetry on $\partial \Lambda$ while the rest of qubits are in the trivial state $|+\rangle$. We then gauge the entire system by coupling it to $\mathbb{Z}_{2}\otimes \mathbb{Z}_{2}$ generalized gauge fields. To construct the output Hilbert space, we place qubits on centers of $cde$-plaquettes and $abe$-plaquettes. The gauged model can be written as $H = H_{up}+H_{down}+H_{wall}$ where $H_{up}$ and $H_{down}$ are identical to those of two copies of the four-dimensional $(2,2)$ toric code while $H_{wall}$ can be viewed as a gapped domain wall inserted along $\partial\Lambda$. 

Recall that the four-dimensional $(2,2)$ toric code has pairs of two-dimensional logical operators and possesses loop-like excitations. Let $e_{1},e_{2}$ and $m_{1},m_{2}$ be loop-like electric charges and loop-like magnetic fluxes in two copies of the four-dimensional toric code. Then one can verify that the domain wall transposes loop-like anyonic excitations as follows:
\begin{align}
m_{a} \rightarrow m_{a}e_{b} ,\qquad m_{b} \rightarrow m_{b}e_{a} ,\qquad 
e_{a} \rightarrow e_{a} ,\qquad e_{b} \rightarrow e_{b}. 
\end{align}
The non-triviality of the domain wall indicates that the underlying SPT wavefunction is non-trivial in the presence of $1$-form symmetry.

\subsection{Two-dimensional model with $0$- and $1$-form symmetry}

So far, we have considered SPT Hamiltonians where charged excitations are either point-like ($0$-form symmetry) or loop-like ($1$-form symmetry). It turns out one can construct a non-trivial SPT Hamiltonian protected by both $0$-form and $1$-form symmetries where the system supports both point-like and loop-like charged excitations. In this subsection, we present a two-dimensional SPT Hamiltonian with $0$ and $1$-form $\mathbb{Z}_{2}\otimes \mathbb{Z}_{2}$ symmetry.

\begin{figure}[htb!]
\centering
\includegraphics[width=0.75\linewidth]{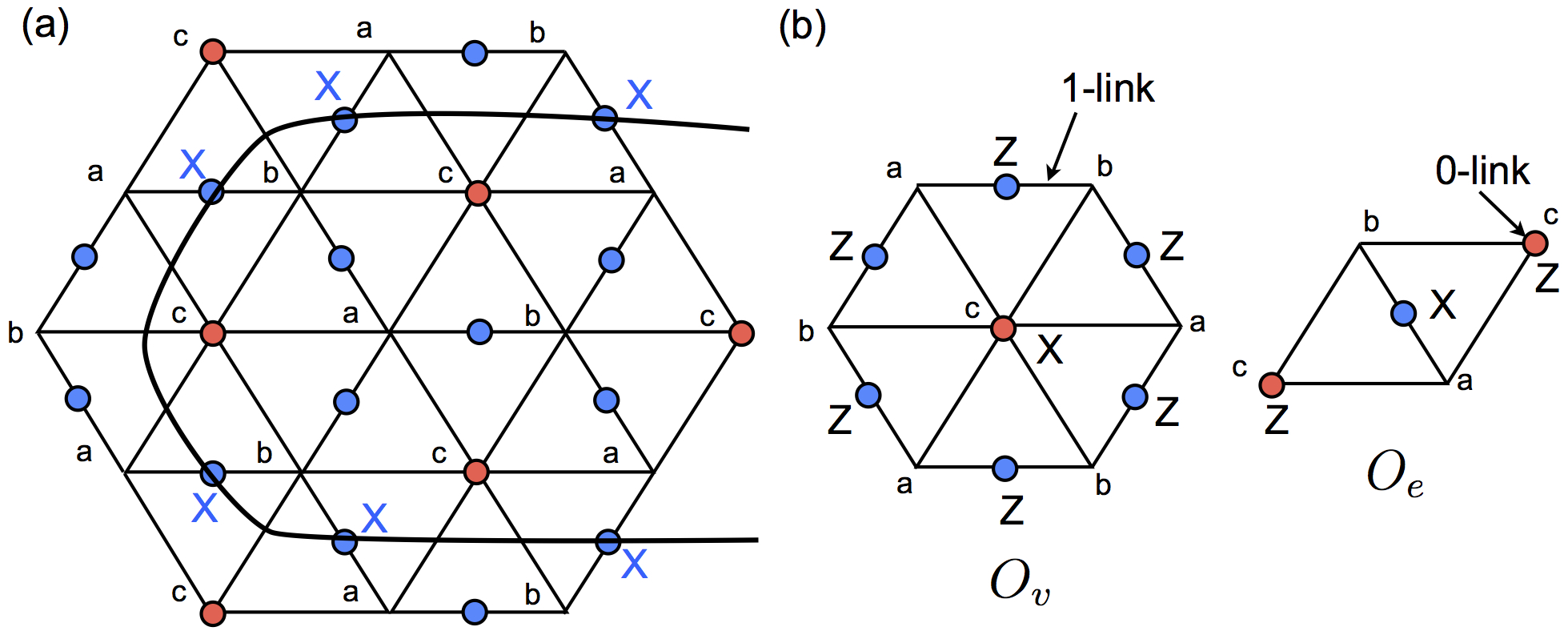}
\caption{SPT phase with $0$- and $1$-form symmetry. (a) A $1$-form symmetry operator which is a closed string of Pauli $X$ operators. (b) Interaction terms. Decorations are on $1$-links and $0$-links.
} 
\label{fig_mixed_form}
\end{figure}

Consider a two-dimensional three-colorable graph with color labels $a,b,c$. We place qubits on $ab$-edges and $c$-vertices as in Fig.~\ref{fig_mixed_form}(a). We start from the trivial Hamiltonian $H_{0}=-\sum_{e\in E^{(ab)}}X_{e} - \sum_{v\in V^{(c)}} X_{v}$ where $E^{(ab)},V^{(c)}$ represent the sets of all the $ab$-edges and $c$-vertices respectively. We shall apply the following finite-depth quantum circuit:
\begin{align}
U^{(0,1)} = \prod_{(i,j)\in \Delta} \mbox{C}Z_{i,j}
\end{align}
where $\mbox{C}Z_{i,j}$ acts on a pair of qubits contained inside a $2$-simplex. The non-trivial SPT Hamiltonian is 
\begin{align}
H_{1}= -\sum_{e\in E^{(ab)}}O_{e} - \sum_{v \in V^{(c)}} O_{v}.
\end{align}
Interaction terms are Pauli $X$ operators with some decorations of Pauli $Z$ operators:
\begin{align}
O_{e} = X_{e}\prod_{v \in \tb{$0$-link}(e) } Z_{v},\qquad 
O_{v} = X_{v}\prod_{e \in \tb{$1$-link}(v) } Z_{E^{(ab)}}
\end{align}
where $\tb{$0$-link}(e),\tb{$1$-link}(v)$ are sets of $0$-links of $e$ and $1$-links of $v$ respectively (see Fig.~\ref{fig_mixed_form}).

The Hamiltonian has $0$-form $\mathbb{Z}_{2}$ symmetry and $1$-form $\mathbb{Z}_{2}$ symmetry acting on $c$-vertices and $ab$-edges respectively. The global $0$-form symmetry operator is given by $S^{(c)} = \prod_{v\in V^{(c)}} X_{v}$ where $V^{(c)}$ represents the set of all the vertices of color $c$. A $1$-form symmetry operator is a string of Pauli $X$ operators $S^{(ab)}(\mathcal{M})$ where $\mathcal{M}$ is a closed loop intersecting with $ab$-edges, $S^{(ab)}(\mathcal{M}) = \prod_{e\in \mathcal{M}} X_{e}$, as shown in Fig.~\ref{fig_mixed_form}(a). To see this, observe that $S^{(ab)}(\mathcal{M})$ can be constructed by taking a product of $O_{e}$ for edges crossed by $\mathcal{M}$ or contained inside $\mathcal{M}$. Excitations associated with violations of $O_{e}$ have geometries of closed loops due to the $1$-form symmetry. 

\begin{figure}[htb!]
\centering
\includegraphics[width=0.80\linewidth]{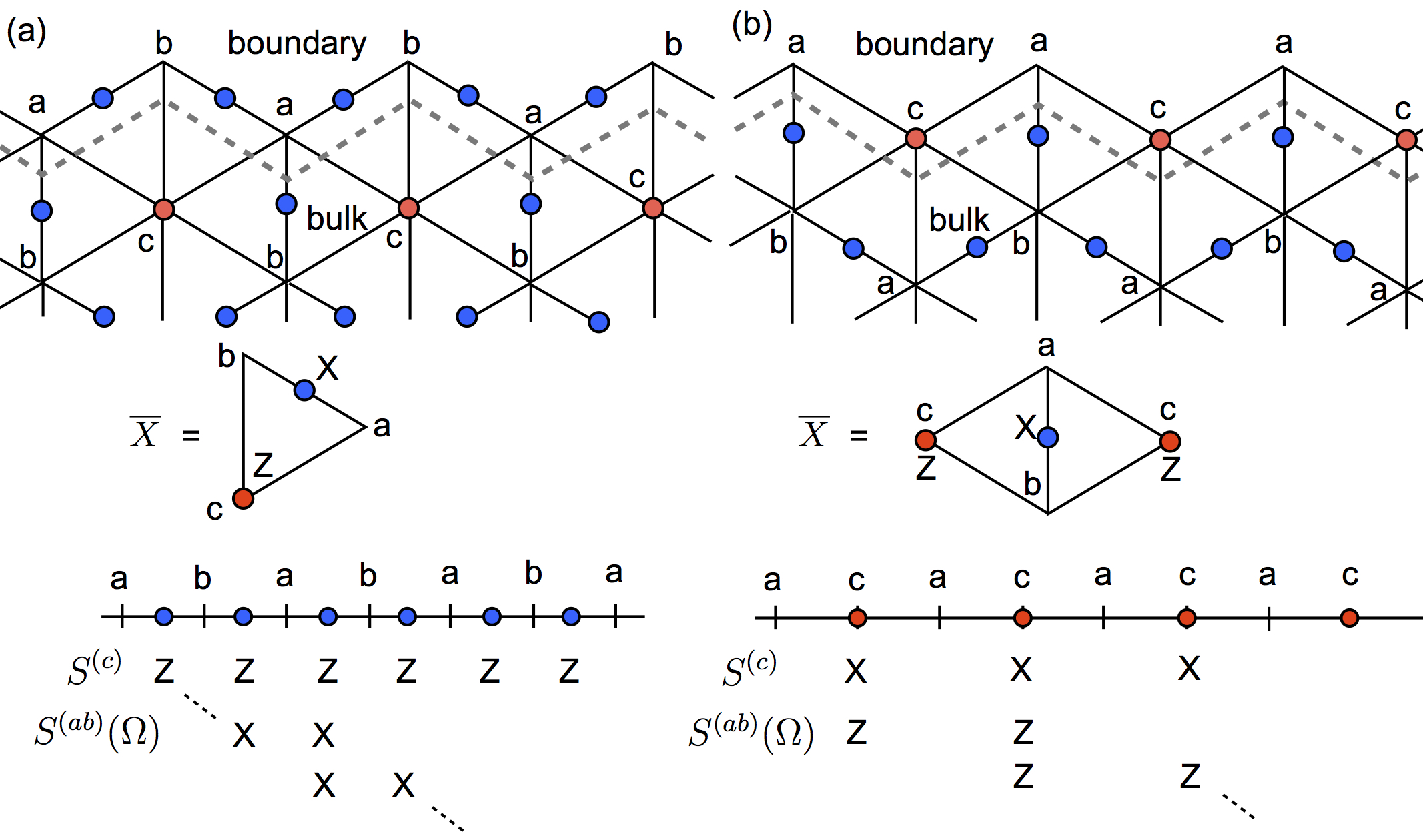}
\caption{(a) A boundary with $a,b$-vertices. (b) A boundary with $a,c$-vertices.
} 
\label{fig_0_1_edge}
\end{figure}

\begin{figure}[htb!]
\centering
\includegraphics[width=0.65\linewidth]{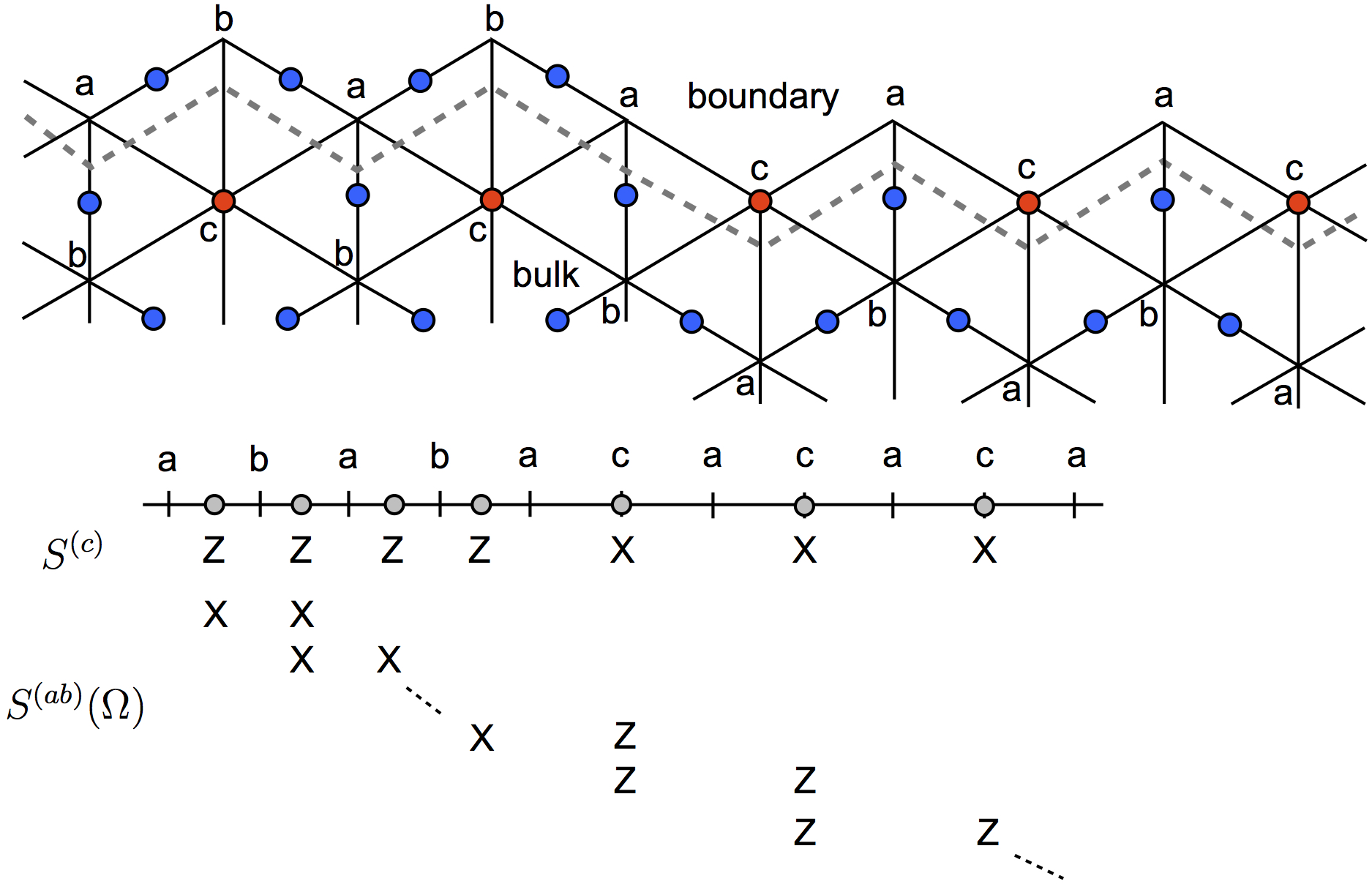}
\caption{A generic boundary consisting of $a,b,c$-vertices. The boundary mode consists of ferromagnets in $X$-basis and $Z$-basis which are smoothly connected.
} 
\label{fig_0_1_edge2}
\end{figure}

To verify the non-triviality of the SPT Hamiltonian, we study its boundary mode. Consider a one-dimensional boundary which consists only of vertices of color $a,b$ as in Fig.~\ref{fig_0_1_edge}(a). See the figure for dressed boundary operators and the action of symmetry operators inside the low-energy subspace. Let us denote $ab$-edges on the boundary by $e_{j}$. Allowed boundary terms are $\overline{X}_{e_{j}}\overline{X}_{e_{j+1}}$ and their products, implying that the boundary mode supports a ferromagnetic order in the $X$ basis. One can also consider a one-dimensional boundary which consists only of vertices of color $a,c$ as in Fig.~\ref{fig_0_1_edge}(b). Let us denote $c$-vertices on the boundary by $v_{j}$. Allowed boundary terms are $\overline{Z}_{v_{j}}\overline{Z}_{v_{j+1}}$ which corresponds to a ferromagnet in the $Z$ basis. Thus, in either choice of boundaries, the boundary mode would be a classical ferromagnet. Finally, one may consider a one-dimensional boundary where the aforementioned two types of boundaries coexist as depicted in Fig.~\ref{fig_0_1_edge2}. See the figure for the action of symmetry operators in the low-energy subspace. The boundary mode, associated with a $\overline{X}_{E_{j}}\overline{X}_{E_{j+1}}$ ferromagnet, is smoothly connect to the boundary mode, associated with a $\overline{Z}_{V_{j}}\overline{Z}_{V_{j+1}}$ ferromagnet. Thus, properties of the boundary mode do not depend on choices of boundaries. 

Next, let us apply the gauging map. Since symmetry operators have different dimensionality, one needs to couple the system to gauge fields of different forms. Specifically, we shall consider the gauging map $\Gamma=\Gamma^{(ab)}\otimes \Gamma^{(c)}$ which is a duality map between the symmetric subspace to the gauge symmetric subspace. Due to the colorability of the graph, the output of $\Gamma^{(ab)}$ lives on $c$-vertices while the output of $\Gamma^{(c)}$ lives on $ab$-vertices. Let us gauge the trivial Hamiltonian $H_{0}$ and non-trivial Hamiltonian $H_{1}$. The gauged model $\hat{H}_{0}$ consists of the toric code living on $ab$-edges and an Ising ferromagnet (the output wavefunction is the GHZ state) living on $c$-vertices. The gauged model $\hat{H}_{1}$ is identical to the original Hamiltonian $H_{1}$ up to transversal Hadamard transformations. Gauged wavefunctions $|\hat{\psi}_{0}\rangle$ and $|\hat{\psi}_{1}\rangle$ are not connected by a finite-depth quantum circuit, so $H_{0}$ and $H_{1}$ belong to different topological phases in the presence of symmetries. 

By gauging this non-trivial two-dimensional wavefunction in three dimensions, one can construct a non-trivial gapped domain wall. The system consists of the $(1,2)$-toric code and the $(2,1)$-toric code with a domain wall. Anyonic excitations are given by $e_{loop},e_{point},m_{loop},m_{point}$. Upon crossing the domain wall, labels of anyonic excitations are transposed as follows:
\begin{align}
e_{loop}\rightarrow e_{loop} \qquad e_{point}\rightarrow e_{point}\qquad  m_{loop}\rightarrow m_{loop}e_{loop}\qquad m_{point}\rightarrow m_{point}e_{point}.
\end{align}

\subsection{Three-dimensional model with $0$-, $0$- and $1$-form symmetry}

In this subsection, we consider a three-dimensional SPT Hamiltonian with $0$-, $0$- and $1$-form $\mathbb{Z}_{2}\otimes \mathbb{Z}_{2}\otimes \mathbb{Z}_{2}$ symmetries. Given a three-dimensional four-colorable graph with color labels $a,b,c,d$, we place qubits on $a$-vertices, $b$-vertices and $cd$-edges. Starting from a trivial Hamiltonian $H_{0}= - \sum_{v\in V^{(a)}}X_{v}- \sum_{v\in V^{(b)}}X_{v}- \sum_{e\in E^{(cd)}}X_{e}$, the non-trivial model is constructed by applying $U^{(0,0,1)}=\prod_{(i,j,k)\in \Delta} \mbox{CC}Z_{i,j,k}$ to the trivial Hamiltonian. Namely,
\begin{align}
H_{1}= - \sum_{v\in V^{(a)},V^{(b)}}O_{v}- \sum_{e\in E^{(cd)}}O_{e}
\end{align}
where $O_{v},O_{e}$ are Pauli $X$ operators with decorations of $\mbox{C}Z$ operators on $2$-links of $v$ and $1$-links of $e$ respectively. The system has three types of $\mathbb{Z}_{2}$ symmetry operators. Global $0$-form symmetry operators are given by $S^{(a)}=\prod_{v\in V^{(a)}} X_{v}, S^{(b)}=\prod_{v\in V^{(b)}} X_{v}$ while a global $1$-form operator is given by $S^{(cd)}(\mathcal{M})=\prod_{e\in \mathcal{M}^{(cd)}} X_{e}$ where $\mathcal{M}^{(cd)}$ is a codimension-$1$ closed manifold intersecting with $cd$-edges. 

\begin{figure}[htb!]
\centering
\includegraphics[width=0.65\linewidth]{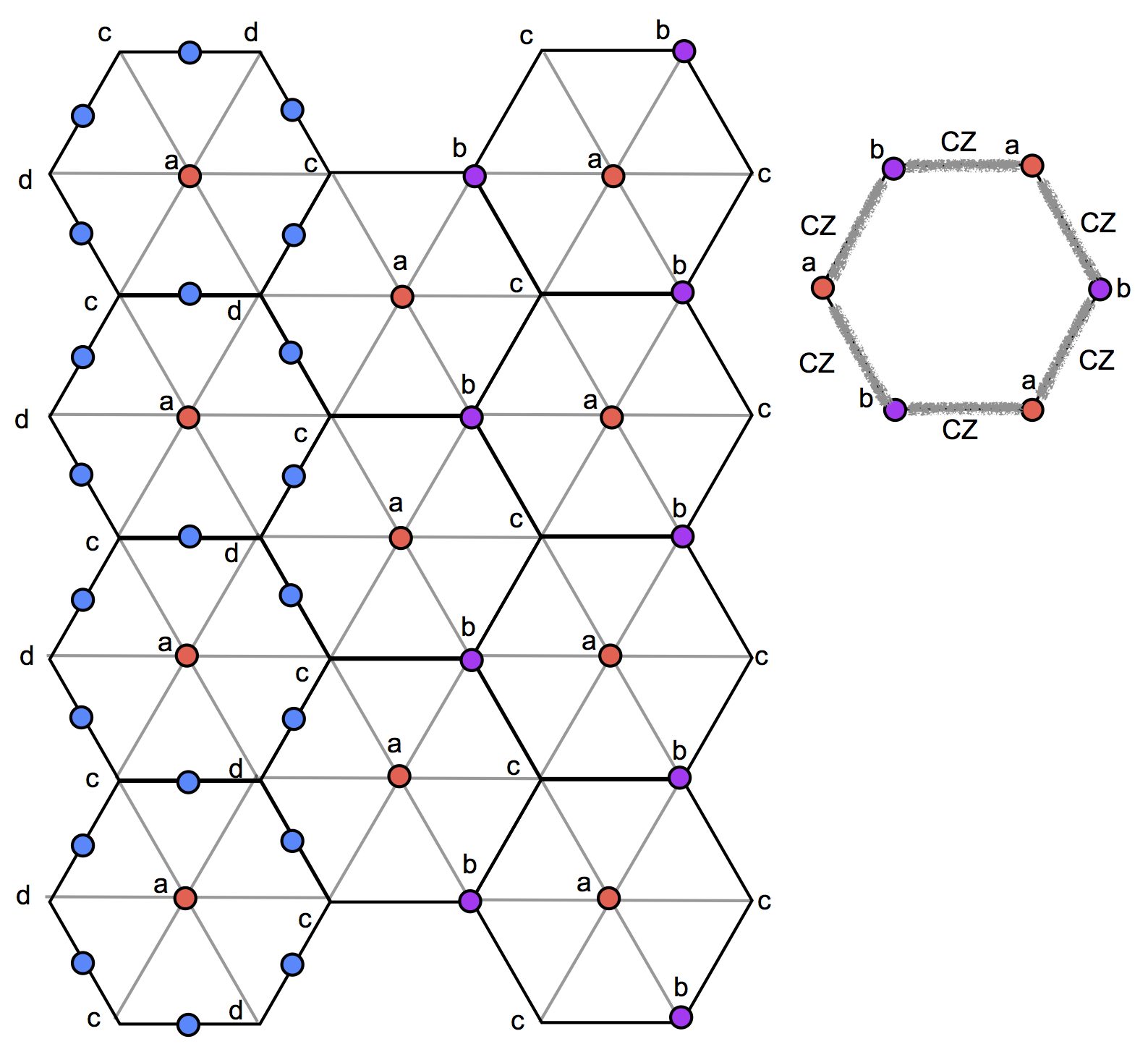}
\caption{Boundary mode for a $(0,0,1)$-form SPT phase. The left side of the boundary consists of vertices of color $a,c,d$ while the right hand side consists of vertices of color $a,b$. A plaquette-like symmetry operator of control-$Z$ operators is also shown. 
} 
\label{fig_001}
\end{figure}

We study the boundary mode of the Hamiltonian. Consider a two-dimensional boundary which consists only of vertices of $a,c,d$ (Fig.~\ref{fig_001}). The boundary can be seen as a three-colorable graph $\partial \Lambda$ with color labels $a,c,d$ where qubits live on $cd$-edges and $a$-vertices. The action of symmetry operators in the low-energy subspace can be written in terms of dressed boundary operators as follows:
\begin{align}
S^{(a)} \sim \prod_{v\in V^{(a)}} \overline{X}_{v} \qquad S^{(cd)}(\mathcal{M}) \sim \prod_{e \in \mathcal{M}} \overline{X}_{e}\qquad S^{(b)} \sim \prod_{(i,j)\in \Delta} \overline{\mbox{C}Z}_{i,j}
\end{align}
where $\mathcal{M}$ is a closed loop on the boundary, $V^{(a)}$ is the set of vertices of color $a$ on the boundary and $\Delta$ is the set of all the $2$-simplex on the boundary graph $\partial \Lambda$. The following quantum critical Hamiltonian commutes with symmetry operators:
\begin{align}
H_{boundary} = - \sum_{v\in V^{(a)}} \overline{X}_{v} - \sum_{e\in E^{(cd)}} \overline{X}_{e} - \sum_{v\in V^{(a)}} \overline{X}_{v}\prod_{ e \in 1\tb{-link}(v)  } \overline{Z}_{e} - \sum_{e\in E^{(cd)}} \overline{X}_{e}\prod_{ v \in 0\tb{-link}(e) }\overline{Z}_{v}.
\end{align}
Note that this Hamiltonian is identical to a summation of trivial and non-trivial two-dimensional SPT Hamiltonians with $0$-form and $1$-form $\mathbb{Z}_{2}\otimes \mathbb{Z}_{2}$ symmetry. This critical Hamiltonian can be transformed into two decoupled copies of critical Ising model by a duality transformation.

Next, let us consider a two-dimensional boundary which consists only of vertices of $a,b,c$ (Fig.~\ref{fig_001}). The boundary graph $\partial \Lambda$ is a three-colorable graph with color labels $a,b,c$ where qubits live on $a$-vertices and $b$-vertices while no qubit lives on $c$-vertices. The actions of symmetry operators are
\begin{align}
S^{(a)} \sim \prod_{v\in V^{(a)}} \overline{X}_{v} \qquad 
S^{(b)} \sim \prod_{v\in V^{(b)}} \overline{X}_{v}\qquad 
S^{(c)}(v^{(c)}) \sim \prod_{(i,j)\in 1\tb{-link}(v^{(c)})  }\overline{\mbox{C}Z}_{i,j}
\end{align}
where $S^{(c)}(v^{(c)})$ is a plaquette-like product of $\mbox{C}Z$ operators as shown in Fig.~\ref{fig_001}. The following quantum critical Hamiltonian, graphically shown, commutes with symmetry operators:
\begin{align}
\includegraphics[height=0.8in]{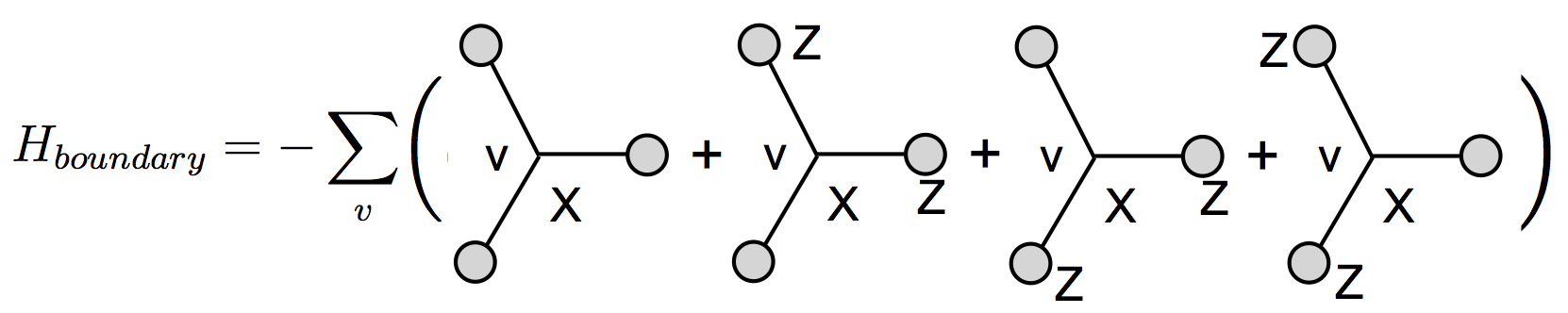}.
\end{align}
With a bit of calculations, one can show that this critical Hamiltonian can be transformed into two decoupled copies of critical Ising model by a duality transformation. Thus, regardless of the choice of boundaries, one obtains two copies of critical Ising model or spontaneous breaking of $\mathbb{Z}_{2}\otimes \mathbb{Z}_{2}$ symmetry. 

By gauging this three-dimensional SPT wavefunction in four dimensions, one can construct a non-trivial domain wall. The system consists of two copies of the $(1,3)$-toric code and one copy of the $(2,2)$-toric code with a domain wall. Let us denote anyonic excitations as follows:
\begin{align}
e^{(1)}_{point},\ e^{(2)}_{point},\ e_{loop}, \
m^{(1)}_{membrane}, \ m^{(2)}_{membrane},\ m_{loop}
\end{align}
where $\{e^{(1)}_{point},m^{(1)}_{membrane}\}, \{e^{(2)}_{point},m^{(2)}_{membrane}\}, \{ e_{loop}, m_{loop} \}$ exhibit non-trivial braiding statistics respectively. The domain wall mixes the above excitations in an intriguing way. Upon crossing the domain wall, electric charges remain unchanged:
\begin{align}
e^{(1)}_{point}\rightarrow e^{(1)}_{point}\qquad e^{(2)}_{point}\rightarrow e^{(2)}_{point} \qquad  e_{loop} \rightarrow e_{loop}
\end{align}
while magnetic fluxes transform into composites of magnetic fluxes and superpositions of electric charges. Namely, one has
\begin{equation}
\begin{split}
&m^{(1)}_{membrane} \rightarrow m^{(1)}_{membrane} s(e^{(2)}_{point},e_{loop}) \qquad m^{(2)}_{membrane} \rightarrow m^{(2)}_{membrane} s(e^{(1)}_{point},e_{loop})\\
&m_{loop} \rightarrow m_{loop} s(e^{(1)}_{point},e^{(2)}_{point})
\end{split}
\end{equation}
where $s(e^{(2)}_{point},e_{loop})$ is a membrane-like object which consists of superpositions of $e^{(2)}_{point}$ and $e_{loop}$, and is characterized by a two-dimensional SPT wavefunction with $0$-form and $1$-form $\mathbb{Z}_{2}\otimes \mathbb{Z}_{2}$ symmetry while $s(e^{(1)}_{point},e^{(2)}_{point})$ is a loop-like object which consists of superpositions of $e^{(1)}_{point}$ and $e^{(2)}_{point}$, and is characterized by a one-dimensional SPT wavefunction with $0$-form $\mathbb{Z}_{2}\otimes \mathbb{Z}_{2}$ symmetry. While the system consists of three copies of the toric code which do not belong to the same topological phase, a gapped domain wall which mixes three copies of the toric code can be constructed. This observation hints rather rich possibilities of gapped boundaries and domain walls in topological phases. 

\subsection{Generic recipe}

Here we summarize the construction of $q$-form SPT Hamiltonians. Consider a $d$-dimensional simplicial lattice which is $d+1$-colorable with color labels $a_{1},\ldots,a_{d+1}$. Consider a partitioning of a positive integer $d+1$ with positive integers $p_{j}$:
\begin{align}
d+1 = p_{1} + p_{2} + \ldots + p_{m},\qquad 1\leq p_{1}\leq p_{2}\leq \ldots \leq p_{m}
\end{align}
Let $q_{j}=p_{j}-1$ and $R_{j}=p_{1}+ \ldots + p_{j-1}$. We will construct a model with $(q_{1},q_{2},\ldots,q_{m})$-form ${\mathbb{Z}_{2}}^{\otimes m}$ symmetry. We place qubits on $q_{j}$-simplices of color $k_{j}:=a_{R_{j-1}+1},\ldots, a_{R_{j}}$. The trivial Hamiltonian is 
\begin{align}
H_{0}= - \sum_{j=1}^{m} \sum_{v^{(q_{j})}\in \Delta^{(q_{j})}} X_{v^{(q_{j})}}
\end{align}
where $\Delta^{(q_{j})}$ represents the set of all the $q_{j}$-simplexes of color $k_{j}$. The non-trivial Hamiltonian $H_{1}$ is constructed by applying the following finite-depth quantum circuit to the trivial Hamiltonian:
\begin{align}
U^{(q_{1},q_{2},\ldots,q_{m})} = \prod_{(i_{1},i_{2},\ldots,i_{m})\in \Delta}{\mbox{C}^{\otimes m-1} Z}_{i_{1},i_{2},\ldots,i_{m}}
\end{align}
where $\Delta$ represents the set of all the $d$-simplexes. The resulting Hamiltonian is
\begin{align}
H_{1}= - \sum_{j=1}^{m} \sum_{v\in \Delta^{(q_{j})}} O_{v^{(q_{j})}}
\end{align}
where 
\begin{align}
O_{v^{(q_{j})}} = X_{v^{(q_{j})}}\prod_{(i_{1},i_{2},\ldots,i_{m-1})\in (d-q_{j}-1)\tb{-link}} \mbox{C}^{\otimes m-2}Z_{i_{1},i_{2},\ldots,i_{m-1}},\qquad v^{(q_{j})}\in \Delta^{(q_{j})}.
\end{align}
A $q_{j}$-form symmetry operator is 
\begin{align}
S^{(k_{j})} (\mathcal{M}^{(q_{j})} )= \prod_{ v^{(q_{j})} \in \mathcal{M}^{(q_{j})} } X_{ v^{(q_{j})}}
\end{align}
where $\mathcal{M}^{(q_{j})}$ is a codimension-$q_{j}$ closed manifold intersecting with $q_{j}$-simplexes. 

When these models are constructed on a manifold with boundaries, non-trivial protected boundary modes can be supported. We do not have complete characterization of boundary modes. For the cases where $m=2$, there are two possible choices of boundary vertices, and boundary modes are either the $(q_{1},d-1-q_{2})$-toric code or the $(q_{2},d-1-q_{1})$-toric code, which are equivalent to each other under Hadamard transformation. So, physics of the boundary mode does not depend on the choices of boundary vertices. For $m>2$, boundaries can support protected gapless modes. Due to topological nature of the theory (\emph{i.e} diffeomorphism invariance), we expect that physics of the boundary mode does not depend crucially on choices of boundaries. However, we do not have independent argument for it.

If one gauges the $d$-dimensional SPT wavefunctions in $d+1$ dimensions, one obtains a gapped domain wall in a $d+1$-dimensional system which consists of the $(p_{j},d+1-p_{j})$-toric code for $j=1,\ldots,m$. The system possesses $q_{j}$-dimensional electric charges, denoted by $e_{j}$, and $(d-1-q_{j})$-dimensional magnetic fluxes, denoted by $m_{j}$. Upon crossing the domain wall, electric charged remain unchanged: $e_{j}\rightarrow e_{j}$ while magnetic fluxes get transformed as follows:
\begin{align}
m_{j}\rightarrow m_{j}s(e_{1},\ldots,e_{j-1},e_{j+1},\ldots,e_{m})
\end{align}
where $s(e_{1},\ldots,e_{j-1},e_{j+1},\ldots,e_{m})$ is a $(d-1-q_{j})$-dimensional superposition of electric charges $e_{1},\ldots,e_{j-1},e_{j+1},\ldots,e_{m}$. Namely, $s(e_{1},\ldots,e_{j-1},e_{j+1},\ldots,e_{m})$ can be characterized by a wavefunction of $(d-1-q_{j})$-dimensional SPT phase with $(q_{1},\ldots,q_{j-1},q_{j+1},\ldots,q_{m})$-form ${\mathbb{Z}_{2}}^{\otimes m-1}$ symmetry. We expect that magnetic fluxes and fluctuating charges will exhibit non-trivial multi-brane braiding statistics.

\section{Fault-tolerant logical gate}\label{sec:gate}

In this section, we comment on applications of our construction to the problem of classifying fault-tolerantly implementable logical gates in topological quantum codes. By topological quantum codes, we mean quantum error-correcting codes, supported on lattices, that can be characterized by geometrically local generators. Namely, we shall argue that all the examples of SPT phases with generalized global symmetry proposed in this paper have corresponding fault-tolerantly implementable logical gates in topological quantum codes living in one more dimensions.

\subsection{Logical gate and domain wall}

We begin by recalling the connection between classifications of fault-tolerant logical gates and gapped domain walls~\cite{Beni15, Beni15c}. The underlying difficulty in quantum information science is the fact that quantum entanglement decays easily and qubits need to be protected from noise and errors. In theory, this challenge can be resolved by using quantum error-correcting codes where single qubit information is encoded in many-body entangled states such that local errors do not destroy the original information. Then, one can perform quantum computation fault-tolerantly inside a protected subspace (codeword space) of a quantum error-correcting code by performing error-correction frequently. 

A naturally arising question concerns how to implement logical gate operations inside the codeword space. Ideally, one hopes to perform logical gates by transversal implementations of unitary operators which have tensor product form, acting on each qubit individually. For such a transversal logical gate, local errors do not propagate to other qubits, and thus its implementation is fault-tolerant. One may also fault-tolerantly implement logical gates which can be expressed as finite-depth local quantum circuits. However, if a logical gate implementation requires a highly non-local and complicated quantum circuit, local errors may propagate to the entire system in a uncontrolled manner. As such, it is important to find/classify fault-tolerantly implementable logical gates in quantum error-correcting codes~\footnote{There are logical gates which do not admit finite-depth circuit implementation, but can be implemented in a rather simple manner. For instance, a Hadamard-like logical gate can be implemented in the two-dimensional toric code by shifting the lattice sites in a diagonal direction, followed by transversal application of Hadamard operators.}. 

To gain some intuition on the restriction on fault-tolerant logical gates, consider the two-dimensional toric code. The system has string-like Pauli $X$ and Pauli $Z$ logical operators which act non-trivially inside the ground state space (codeword space) and have transversal form. However, the toric code does not admit any other transversal logical gates except for products of string-like Pauli operators. In this sense, the toric code has a rather restricted set of transversal logical gates. Thus, we hope to find quantum codes with a larger set of fault-tolerant logical gates~\footnote{However, a larger set of fault-tolerant logical gates often implies weaker error tolerance. See~\cite{Pastawski15}}.

\begin{figure}[htb!]
\centering
\includegraphics[width=0.65\linewidth]{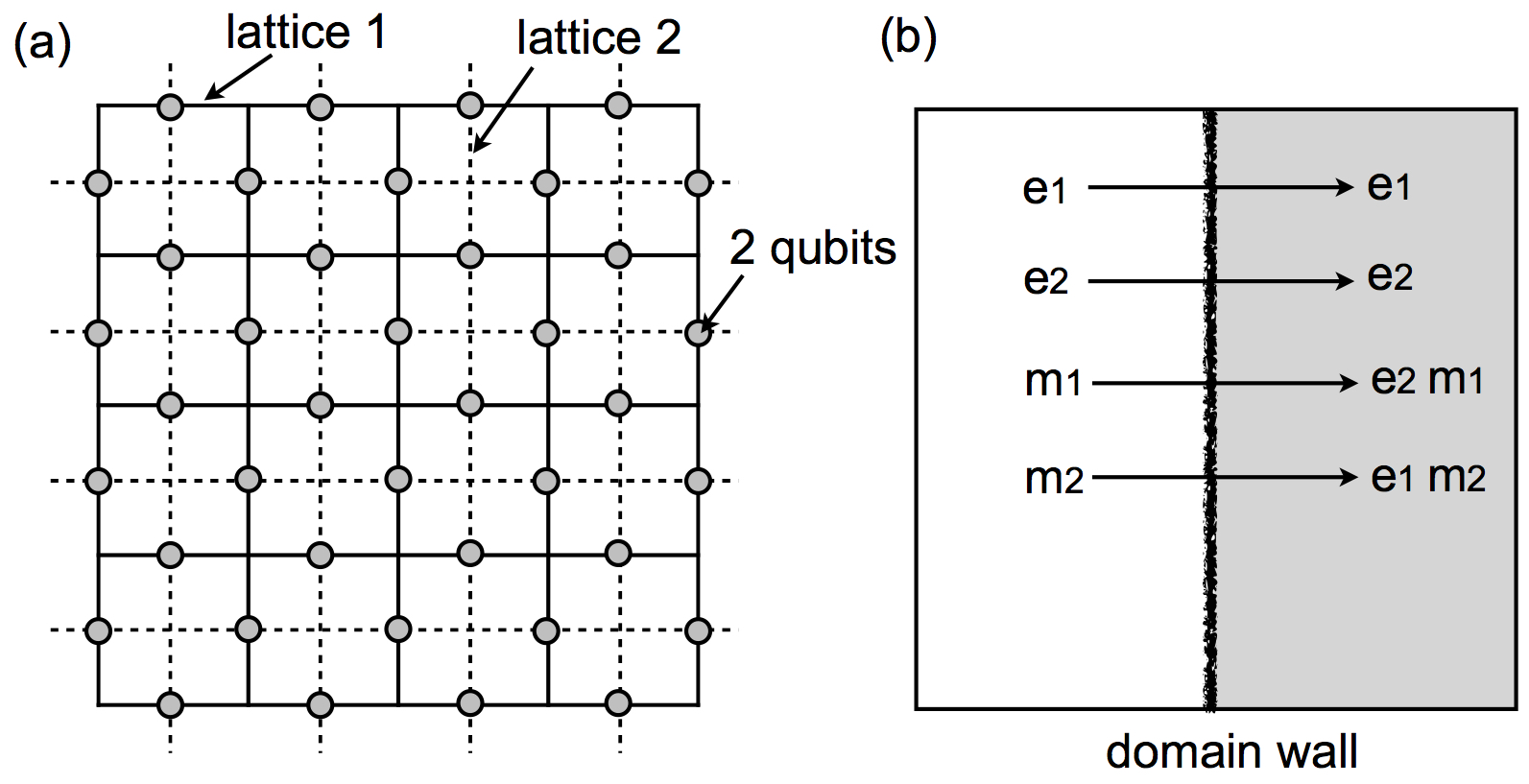}
\caption{(a) Two copies of the toric code. The first copy lives on a square lattice with solid lines while the second copy lives on a dual lattice with dotted lines. Grey dots represent two qubits from each copy of the toric code. Control-$Z$ gates are applied to pairs of qubits at the same sites. (b) A gapped domain wall constructed from a control-$Z$ logical gate. We apply the logical gate only on the right hand side of the lattice. 
} 
\label{fig_domain_wall}
\end{figure}

Let us look at an example where one can implement non-Pauli logical operators transversally. Consider a two-dimensional system which consists of two decoupled copies of the toric code. Namely, we assume that the first copy lives on a square lattice and the second copy lives on a dual lattice as shown in Fig.~\ref{fig_domain_wall}(a). The Hamiltonian is given by
\begin{align}
H = - \sum_{v\in V^{(1)}} A_{v}^{(1)} - \sum_{v\in V^{(2)}} A_{v}^{(2)} - \sum_{p\in P^{(1)}} B_{p}^{(1)} - \sum_{p\in P^{(2)}} B_{p}^{(2)}
\end{align}
where two qubits are placed at each site. Here $V^{(1)},V^{(2)}$ represent the sets of vertices on lattice $1$ and $2$ respectively while $P^{(1)},P^{(2)}$ represent the sets of plaquettes. In this system, one can apply a logical control-$Z$ gate among two copies of the toric code transversally. Specifically, we apply control-$Z$ gates to pairs of qubits in the first and second copy at the same sites. This transversal operation preserves the ground state space, but has non-trivial action on it. Recalling the transformation rules under conjugations by control-$Z$ gates, $\mbox{C}Z (X_{1}) \mbox{C}Z= (X_{1})Z_{2},\mbox{C}Z (X_{2}) \mbox{C}Z= Z_{1}(X_{2})$, one notices that transversal control-$Z$ gates transform anyonic excitations as follows:
\begin{align}
e_{1}\rightarrow e_{1} \qquad m_{1} \rightarrow e_{2}m_{1}\qquad
e_{2}\rightarrow e_{2} \qquad m_{2} \rightarrow e_{1}m_{2}.\label{eq:2dim_CZ}
\end{align}
In general, membrane-like fault-tolerant logical operators transpose labels of anyonic excitations in two-dimensional topological quantum codes~\cite{Bravyi13b}.

In order to construct a gapped domain wall, let us split the entire system into the left and right parts and apply the transversal control-$Z$ gate only on the right hand side of the lattice. This transforms the Hamiltonian into the following form:
\begin{align}
H = H_{left} + H_{wall} + H_{right}
\end{align}
where $H_{left}$ and $H_{right}$ remain unchanged while $H_{wall}$ can be viewed as a gapped domain wall which connects $H_{left}$ and $H_{right}$. Upon crossing the domain wall, anyonic excitations are transposed according to Eq.~(\ref{eq:2dim_CZ}). As this observation implies, given a $d$-dimensional non-trivial fault-tolerant logical gates in a $d$-dimensional topological quantum code, one can construct a corresponding domain wall since non-trivial logical gates would transform types of excitations. 

\subsection{Logical gates and generalized global symmetry}

Next, let us establish the connection between SPT phases and gapped domain walls in the context of fault-tolerant logical gates. For this purpose, we revisit the gapped domain wall in the two-dimensional $\mathbb{Z}_{2}\otimes \mathbb{Z}_{2}$ toric code. Consider a two-dimensional trivial symmetric system where a one-dimensional SPT wavefunction with $\mathbb{Z}_{2}\otimes \mathbb{Z}_{2}$ symmetry is inserted as in Fig.~\ref{fig_0_1_edge}(a). By gauging the entire system with respect to $\mathbb{Z}_{2}\otimes \mathbb{Z}_{2}$ symmetry, one obtains two copies of the toric code with a gapped domain wall as shown in Fig.~\ref{fig_0_1_edge}(b). Let $\gamma$ be a one-dimensional line where a one-dimensional SPT wavefunction is placed. Observe that one can move one-dimensional SPT wavefunction to a different location $\gamma'$ by applying a symmetric local unitary transformation $U$ which acts only on qubits enclosed by $\gamma$ and $\gamma'$. By gauging the system, this process of moving one-dimensional SPT wavefunction is equivalent to moving a domain wall from $\gamma$ to $\gamma'$ by applying a local unitary transformation $\hat{U}$ on qubits enclosed by $\gamma$ and $\gamma'$. By sweeping the domain wall over the entire system, one can implement a non-trivial logical gate. Namely, this implements the control-$Z$ gate among two copies of the toric code. In this sense, a one-dimensional SPT wavefunction with $\mathbb{Z}_{2}\otimes \mathbb{Z}_{2}$ symmetry corresponds to the control-$Z$ logical gate in two copies of the two-dimensional toric code. Here we would like to emphasize that the logical gate acts on a system with intrinsic topological order (the $\mathbb{Z}_{2}\otimes \mathbb{Z}_{2}$ toric code) while the logical gate was constructed by gauging the one-dimensional $\mathbb{Z}_{2}\otimes \mathbb{Z}_{2}$ SPT wavefunction. 

In~\cite{Beni15c}, we employed this idea to construct fault-tolerantly implementable logical gates in the $d$-dimensional quantum double model. Namely, for the $d$-dimensional quantum double model with finite group $G$, the domain wall can be constructed by gauging a $(d-1)$-dimensional SPT wavefunction in $d$ dimensions. In other words, one is able to construct logical gates, gapped boundaries and domain walls by using $d$-cocyle functions in the $d$-dimensional quantum double model. We then demonstrated that, for non-trivial domain walls, there exist corresponding non-trivial logical gates which can be implemented by finite-depth local quantum circuits. It was also found that the non-triviality of domain walls can be verified by computing slant products sequentially.

\begin{figure}[htb!]
\centering
\includegraphics[width=0.65\linewidth]{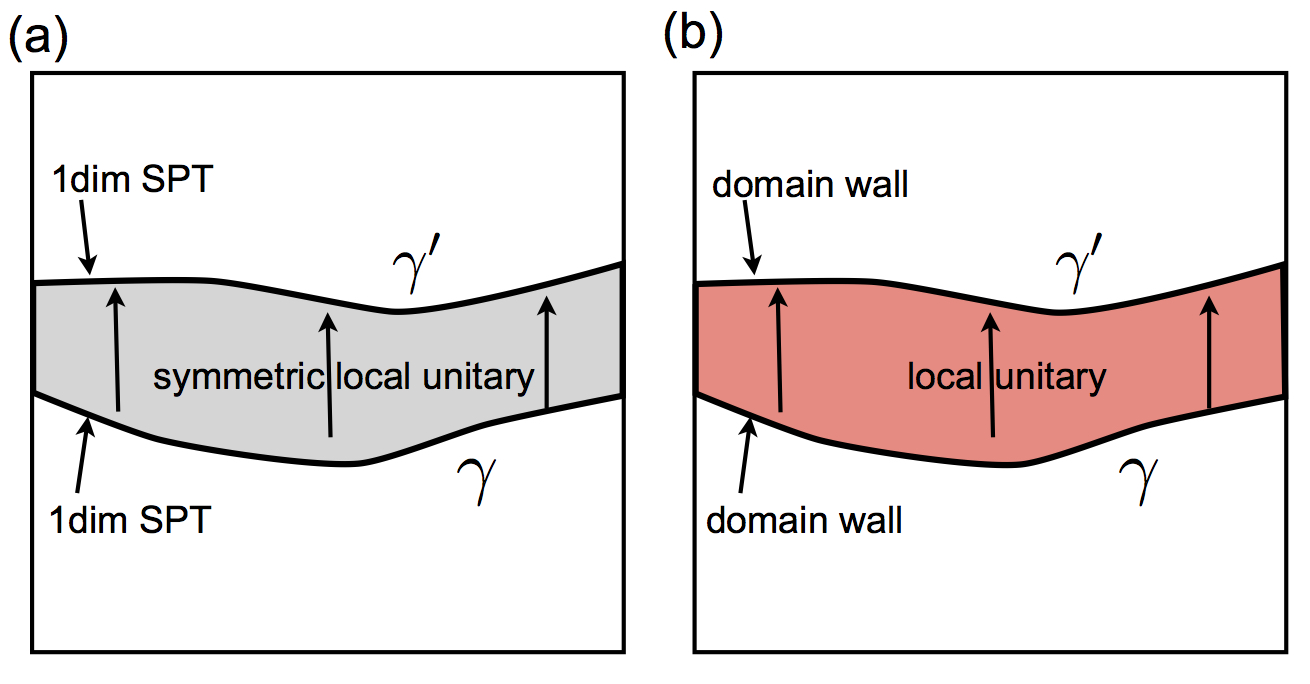}
\caption{A fault-tolerant logical gate constructed from SPT wavefunctions and gapped domain walls. (a) Moving a one-dimensional SPT wavefunction by finite-depth symmetric quantum circuits. (b) Moving a gapped domain wall by finite-depth quantum circuits.
} 
\label{fig_0_1_edge}
\end{figure}

A similar argument applies to SPT phases with generalized global symmetry. Consider a $d$-dimensional SPT phase with $(q_{1},\ldots,q_{m})$-form $(\mathbb{Z}_{2})^{\otimes m}$ symmetry. By gauging this SPT phase in $d+1$ dimensions, one obtains a gapped domain wall in a $d+1$-dimensional topological phase. This $d+1$-dimensional system consists of $(p_{j},d+1-p_{j})$-toric code for $j=1,\ldots,m$. One can move $(d-1)$-dimensional SPT wavefunctions by applying symmetric finite-depth quantum circuits. This implies that one can sweep the domain wall over the entire system by applying a finite-depth local unitary circuit. With some speculation, one notices that this implements a $\mbox{C}^{\otimes m-1}Z$ logical gate among $(p_{j},d+1-p_{j})$-toric code. For instance, by using a three-dimensional SPT phase with $(0,0,1)$-form symmetry, one can construct a $\mbox{C}^{\otimes 2}Z$ logical gate acting among two copies of the $(1,3)$-toric code and one copy of the $(2,2)$-toric code in four dimensions. It is interesting to observe that logical gates can be implemented among several copies of the toric code which belong to different quantum phases. Our findings hint rich possibilities of fault-tolerant logical gates, as well as gapped domain walls, in topological phases of matter. 

\section{Discussions}\label{sec:discussion}


We will conclude the paper with discussions on generalizations of the model and implications of our results. 

\subsection{Generalizations}

While our treatment has been limited to systems with $\mathbb{Z}_{2}$ symmetries, the construction can be generalized to systems with arbitrary abelian symmetries. Here we illustrate the idea for $\mathbb{Z}_{N}$ symmetry. We define the following generalized $m$-qudit control-$Z$ gate
\begin{align}
U^{(m)}=\sum_{g_{1},\ldots,g_{m}}\exp\left(i\frac{2\pi}{N}g_{1}\cdots g_{m}\right)|g_{1},\ldots,g_{m}\rangle\langle g_{1},\ldots,g_{m}|
\end{align}
where $g_{j}=0,\ldots,N-1$. For $N=2$, this reduces to the multi-qubit control-$Z$ gate. To construct SPT phases with $0$-form symmetries, consider a $(d+1)$ colorable graph $\Lambda$ in $d$ dimensions and assign qudits ($N$-state spins) to vertices. On a colorable graph, one is able to assign parity $P(\Delta)=\pm1$ to each $d$-simplex such that neighboring simplexes have opposite parity signs~\cite{Kubica15}. We shall apply the following local unitary to the trivial symmetric Hamiltonian
\begin{align}
U = \prod_{\Delta} (U^{(d+1)}_{\Delta})^{P(\Delta)}.
\end{align}
where $U^{(d+1)}_{\Delta}$ acts on $d+1$ qudits on the $d$-simplex $\Delta$. The system possesses $d+1$ copies of $\mathbb{Z}_{N}$ symmetries, associated with each different color label $a_{1},\ldots,a_{d+1}$. To construct SPT phases with higher-form $\mathbb{Z}_{N}$ symmetries, one places qudits according to the partition of color labels.

Our construction of SPT phases does not exhaust all the possible bosonic SPT phases. Yet, by changing the choices of symmetry operators, one is able to construct some other SPT phases. Let us illustrate the idea by considering two-dimensional SPT phases with $\mathbb{Z}_{2}$ symmetry. The proposed model possesses three copies of $\mathbb{Z}_{2}$ symmetries, captured by three symmetry operators $S_{A}\otimes S_{B}\otimes S_{C}$, associated with three color labels $A,B,C$. It is possible to view the model as an SPT Hamiltonian with one copy of $\mathbb{Z}_{2}$ symmetry by imposing $S_{A}S_{B}S_{C}$ as the single $\mathbb{Z}_{2}$ symmetry operator. This reduces the model to the one proposed by Levin and Gu~\cite{Levin12}. One may also choose to impose two copies of $\mathbb{Z}_{2}$ symmetries by using $S_{A}\otimes S_{B}S_{C}$. Thus, there are three possible SPT phases associated with symmetries:
\begin{align}
S_{A}S_{B}S_{C} \qquad S_{A}\otimes S_{B}S_{C} \qquad S_{A}\otimes S_{B}\otimes S_{C}. 
\end{align}
Recall that, in two spatial dimensions, SPT phases protected by abelian symmetries can be classified into three classes, call type-I, type-II and type-III~\cite{Propitius95}. Levin and Gu showed that the $S_{A}S_{B}S_{C}$ model corresponds to the type-I model. In~\cite{Beni15c}, we showed that the $S_{A}\otimes S_{B}\otimes S_{C}$ model corresponds to the type-III model by the domain wall argument. It is an interesting question to determine the type of the $S_{A}\otimes S_{B}S_{C}$ model.

\subsection{Previously known models}

We comment on the relations between our models and previously known models in three dimensions. Our model can be characterized by partitions of the integer $4$ (which is the space-time dimension). Namely, possible models are characterized by $(1,1,1,1)$, $(2,1,1)$, $(2,2)$ and $(3,1)$. The $(1,1,1,1)$ model possesses $0$-form SPT order with $S_{A}\otimes S_{B}\otimes S_{C}\otimes S_{D}$. One can pick different choices of symmetries, such as $S_{A}S_{B}\otimes S_{C}\otimes S_{D}$, to construct different classes of $0$-form SPT phases. The $(2,2)$ model possesses $1$-form symmetries only, and seems to be described by a theory containing $B\wedge B$ term after gauging. Such TQFTs typically belong to a (rather simple) subclass of the Walker-Wang model~\cite{Walker11}. The $(3,1)$ model involves $2$-form symmetries which have not been discussed much in the literature. However, we think that the model can be reduced to a known one by a duality transformation which exchanges charges and fluxes (in other words, Pauli $X$ and $Z$ operators). The $(2,1,1)$ model has one copy of $1$-form symmetry and two copies of $0$-form symmetries, denoted by $S_{AB}\otimes S_{C}\otimes S_{D}$. Consider a realization with $S_{AB}\otimes S_{C}S_{D}$. After gauging, the model seems to be identical to the Birmingham-Rakowski model with $\mathbb{Z}_{2}$ symmetry introduced in~\cite{Birmingham95}. The Birmingham-Rakowski model can be discussed in a more generic framework, called the Mackaay TQFT where degrees of freedom are placed both on edges and faces as in our construction~\cite{Mackaay00}. As for the model with $S_{AB}\otimes S_{C}\otimes S_{D}$ symmetry, we were not able to find similar constructions in the literature. Whether this construction is truly new or not waits further verifications.

Kapustin and Thorngren utilized the notion of $2$-group to construct SPT phases protected by $0$-form and $1$-form symmetries on lattices.  An important difference between our approach and the $2$-group construction is that the $2$-group construction possesses non-flat connections while our models consist only of flat connections. In simpler words, our models modify $X$-type vertex terms only while the $2$-group construction modifies $Z$-type plaquette terms too. However, duality transformations often allow one to construct models with non-flat connections from the ones with flat connections. Indeed, some of the $2$-group constructions are equivalent to models only with flat connections. Similarly, some of our models are equivalent to models with non-flat connections via duality transformations. Such complementary viewpoints may allow one to construct further examples of interesting TQFT models. At this moment, we were not able to fit our $(2,1,1)$ model into the $2$-group construction despite the fact that the model has $0$-form and $1$-form symmetries only. In~\cite{Kapustin14}, Kapustin and Thorngren briefly mention possible generalizations using the notion of $q$-group ($q>2$). Presumably, such constructions will involve several symmetry operators of different dimensionality, and our construction may give concrete examples of such generalizations.



\subsection{Future problems}

Other questions and future problems are listed below. We did not discuss physical properties of gauged models in depth. Upon gauging SPT Hamiltonians, one typically obtains topologically ordered Hamiltonians whose braiding statistics are twisted due to the decorations added to matter fields. We expect that gauged versions of our models will exhibit rather exotic topological order which may be beyond known theoretical frameworks. For one thing, in two dimensions, the $0$-form $\mathbb{Z}_{2}\otimes \mathbb{Z}_{2}\otimes \mathbb{Z}_{2}$ SPT phase considered in this paper has been shown to be dual to a non-abelian topological phase upon gauging~\cite{Propitius95, Beni15, Beni15c}. This hints a possibility of interesting non-abelian statistics which involve both particles and loops by gauging higher-form SPT phases. SPT phases with $q$-form global symmetry provide a number of interesting quantum critical Hamiltonians as boundary modes. Analytical and numerical studies of such boundary modes may provide further insights into problems of quantum criticality in higher dimensions. Spatial dimension of symmetry operators can be non-integer~\cite{Haah11, Beni13}. Namely, one can construct an SPT Hamiltonian protected by fractal-like symmetry operators. Studies of such fractal SPT phases and their gauged models may be an interesting future problem with applications to efficient magic state distillations. One drawback of our approach is that the proposed models do not have full diffeomorphism invariance due to the use of colorable graphs. While it is possible to coarse-grain or fine-grain graphs by retaining colorability~\cite{Kubica15b}, the full verification of fixed-point properties is an important future problem. Finding field theoretical descriptions of proposed model is also an interesting project~\cite{J_Wang15b}.


\section*{Acknowledgment}

I would like to thank Zhengcheng Gu, Aleksander Kubica, John Preskill, Burak \c{S}ahino\u{g}lu, Ryan Thorngren, Michael Walter and Dominic Williamson for helpful discussions, conversations and/or comments. Part of the work was completed during the visit to the Kavli Institute for Theoretical Physics. We acknowledge funding provided by the Institute for Quantum Information and Matter, an NSF Physics Frontiers Center with support of the Gordon and Betty Moore Foundation (Grants No. PHY-0803371 and PHY-1125565). I am supported by the David and Ellen Lee Postdoctoral fellowship.
This research was supported in part by the National Science Foundation under Grant No. NSF PHY11-25915.


\end{document}